\documentclass[journal]{IEEEtran}

\usepackage{amsmath,amssymb,amsfonts}
\usepackage{graphicx}
\usepackage{svg}
\usepackage{cite}
\usepackage[hidelinks]{hyperref}
\usepackage{xcolor}
\usepackage{stfloats}
\usepackage[caption=false]{subfig}

\newtheorem{remark}{Remark}[section]
\newtheorem{assumption}{Assumption}[section]

\usepackage{cleveref}
\crefname{equation}{}{}
\Crefname{equation}{Equation}{Equations}
\crefname{figure}{Fig.}{Figs.}
\crefname{table}{Table}{Tables}
\crefname{section}{Section}{Sections}

\begin{document}

\title{Robust Tuning of Model Predictive Control for MMC-Based High-Voltage Power Systems}

\author{Victor~Daniel~Reyes~Dreke,~\IEEEmembership{Member,~IEEE,}
        Rahul~Rane,~\IEEEmembership{Member,~IEEE,} \\
        and~Aleksandra~Leki\'{c},~\IEEEmembership{Senior Member,~IEEE}% <-this % stops a space
\thanks{All the authors are associated with Intelligent Electrical Power Grids group, Electrical Sustainable Energy department, Delft University of Technology, Netherlands, e-mail:v.d.reyesdreke@tudelft.com.}% <-this % stops 

\thanks{The work is supported by Horizon Europe project PROSECCO, under grant agreement 101160687, and by NWO Veni project SAFE-GRID, with project number 20248.}}

% \markboth{Journal of \LaTeX\ Class Files,~Vol.~14, No.~8, August~2015}%
% {Shell \MakeLowercase{\textit{et al.}}: Bare Demo of IEEEtran.cls for IEEE Journals}

\maketitle

\begin{abstract}
High-voltage direct current (HDVC) transmission systems based on modular multilevel converters (MMCs) have become a key topology in modern power systems. 
The dynamics of MMCs exhibit strong multivariable coupling, constraints, and uncertainties, motivating the use of model predictive control (MPC) to enhance current regulation performance. 
However, MPC tuning is nontrivial and does not inherently guarantee stability or robustness, particularly in the presence of model uncertainties.
This paper proposes a MPC tuning method that ensures robust performance under bounded model uncertainties. 
This method solves a convex linear optimization problem to compute the optimal weighting matrices $Q$, $R$, and $P$ ensuring optimality and reproducibility. 
As a result,  robustness is enhanced without increasing the online computation burden.
The effectiveness of the method is validated through testing on a real-time digital simulator (RTDS) model of a point-to-point HVDC  system. 
Results demonstrate improved performance compared to conventional LQR-based MPC tuning.
\end{abstract}

% Note that keywords are not normally used for peer review papers.

% \fi
\IEEEpeerreviewmaketitle

% ============================================================================ %
\section{Introduction}
\label{sec:Introduction}
% ============================================================================ %

Modular multilevel converter (MMC)–based high-voltage direct current (HVDC) systems have become a key topology in modern power systems due to their capability to facilitate the integration of renewable energy resources \cite{Perez2012_MMC_Literature_review}. 
However, MMC dynamics are characterized by strong multivariable coupling, constraints, and uncertainties \cite{Wang2019_MMC_Model_Unstruct}. 
These features pose significant challenges in the design of MMC control algorithms, as they can degrade control performance if not explicitly accounted for \cite{Louybary2024_MMC_LMI_Uncertainties}. 
Consequently, there is a need for reliable control strategies capable of regulating MMC currents in the presence of such phenomena~\cite{Yao2025_MMCUncertanties_Struc}.

Recent MMC research has largely focused on two main approaches to improve control performance: (\emph{i}) robust control and (\emph{ii}) model predictive control. 
Robust control techniques are well-suited to address multivariable coupling and both structured and unstructured uncertainties. 
In this context, existing solutions commonly rely on design methods such as $\mathcal{H}_{\infty}$ synthesis \cite{Tavakoli2021_OptimalMMC,Gil2017_RobustHVDC}, $\mu$-synthesis \cite{Tavakoli2022_RobustHVDC}, single value decomposition (SVD)-based performance analysis \cite{Dejene2021_RobustHVDC}, and linear matrix inequalities (LMIs)-based optimal robust control synthesis \cite{Belhaouane2019_RobustHVDC,Ayari2017_RobustHVDC,Olalla2009_MMC_LMI_Robust_Uncertainties}.
These methods yield state or output feedback controllers that ensure robust performance in the presence of disturbances and model uncertainties arising from load and PLL errors \cite{Gil2017_RobustHVDC}, components tolerances \cite{Belhaouane2019_RobustHVDC,Ayari2017_RobustHVDC}, grid impedance variations \cite{Tavakoli2022_RobustHVDC}, and unmodeled dynamics~\cite{Olalla2009_MMC_LMI_Robust_Uncertainties}.
Although these approaches ensure disturbance attenuation and global robust stability, they do not guarantee constraint satisfaction. 
In this case, saturation blocks and current limiters are needed to prevent safety and performance violations, such as overcurrents or overmodulation~\cite{Bergna2012_MMC_Energy}. 
However, adding these elements may compromise closed-loop asymptotic stability \cite{Tarbouriech2011_constraint_stability} and complicate theoretical analysis, particularly in reference-tracking scenarios \cite{Khalil2002nonlinear}.

Model predictive control (MPC) is a promising alternative, as it computes optimal control actions while explicitly enforcing system constraints. 
The MPC capability to enhance MMC current control loop performance has been demonstrated in  \cite{Dekka2019_MPCMMCSurvey,Reyes2022_NMPCHVDC,Shetgaonkar2023_MPCHVDC,Zhang2020_MMCHVDCSurvey}. 
However, its high computational burden,  lack of stability guarantees, or sensitivity to uncertainties are common challenges that hinder a wider use  \cite{Dekka2019_MPCMMCSurvey}. 
Achieving the desired level of robustness with classical MPC requires tuning prediction horizons, weighting matrices, and constraint sets, a process that is often heuristic and time-consuming in practice. Alternative MPC formulations, such as tube-based MPC \cite{Mayne2005_RobustMPC}, min–max MPC \cite{Raimondo2009_RobustMPC}, adaptive MPC \cite{kim2010_Adpativerecent}, and stochastic MPC \cite{Mesbah2016_stochasticMPC}, provide a priori robustness guarantees.
Many of these methods have been applied to converter-level current regulation see e.g. \cite{Heydari2022_Adaptive,kiani2024_MPClearning,tregubov2023_MPCrobust,dragivcevic2017_MPCmodel}; however, their design complexity and computational burden are significantly higher than those of classical MPC~\cite{babayomi2025robust}. 
Consequently, their use has been largely limited to two-level converters with a prediction horizon of one, typically using single-input single-output models suitable for finite-control-set MPC~\cite{babayomi2025robust}. 
This trade-off helps explain why, to the best of the authors’ knowledge, there are not reported uses of robust MPCs in real-time HVDC systems.
% set-based MPC \cite{De2009_RMPC_Set_Baseddesign}

This paper addresses the gap between robust control design and MPC implementation for MMC-based HVDC systems. 
An automatic tuning method is proposed to compute the MPC weighting matrices by leveraging the matching-control approach in \cite{ReyesDreke_DBMC_2023}.  
Unlike \cite{ReyesDreke_DBMC_2023}, we propose an LMI method to compute a robust yet less conservative matching controller. 
The resulting MPC inherits robustness properties without increasing the online computational burden. 
The optimal tuning via matching-control is formulated as a convex linear optimization problem, ensuring global optimality and systematic design.
The approach is validated using simulations on real-time digital platforms (RTDS\textsuperscript{\textregistered}). 
Compared to conventional LQR-based tuning, the proposed method guarantees closed-loop stability and robustness performance, enabling direct enforcement of phase and gain margin requirements specified by TSOs under parametric uncertainty.

% The paper is organized as follows. Section~\ref{sec:MMC_Modeling} presents the MMC model. Section~\ref{sec:RMPC_Design} describes the RMPC design. Section~\ref{sec:Study_Case} introduces the case study, and Section~\ref{sec:Simulation_results} discusses the RTDS simulation results.

% ============================================================================ %
\section{MMC Modeling}
\label{sec:MMC_Modeling}
% ============================================================================ %

\begin{figure}[!hbtp]
    \centering
    \includegraphics[width=\linewidth]{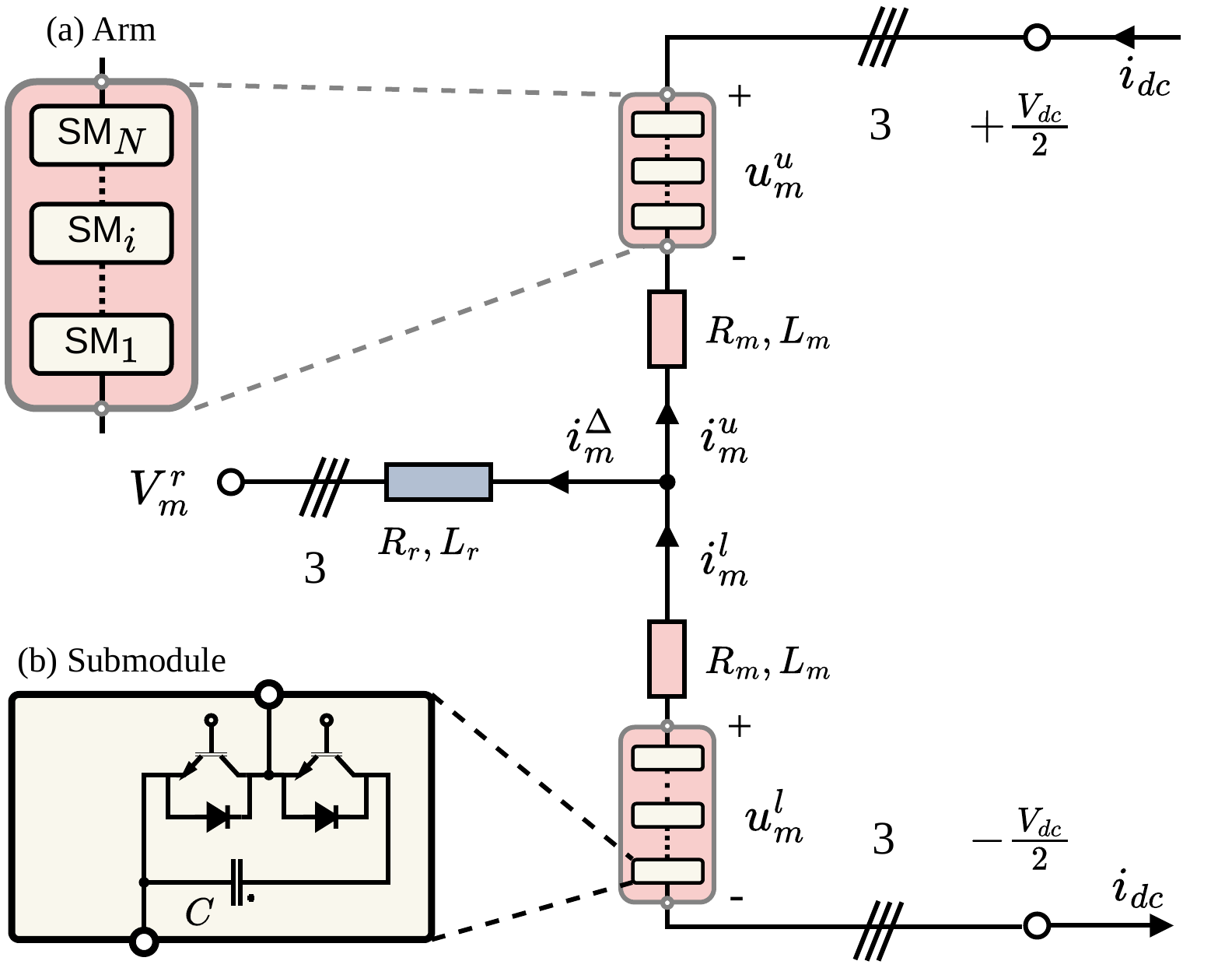}
    \caption{Simplified MMC circuit per phase $m \in \{a,b,c\}$.}
    \label{fig:Diagram_MMC_Average_Circuit}
\end{figure}

MMCs used in HVDC transmission systems are three-phase rectifier/inverter, comprising 3 legs, one for each phase $m\in \{a,b,c\}$. Each leg $m$ consists of two arms $n \in \{u,l\}$, upper and lower arm, having $N_s$ half-bridge submodules (SMs) connected in series to an arm resistor ($R_m$) and inductor ($L_m$). 
The midpoint of each MMC leg $m$ is connected to the transformer resistor ($R_r$) and inductor ($L_r$).
\cref{fig:Diagram_MMC_Average_Circuit} illustrates a simplified diagram of a single-phase MMC circuit in the $abc$-reference framework, where the SMs are the blocks SM$_{i}$. 

The MMC power transfer is determined by the output and circulating currents, which are defined as
\begin{equation}
     i^{\Delta}_{m}  =  i^u_{m}- i^l_{m}, \quad \text{and} \quad i^{\Sigma}_{m}  = (i^u_{m}+ i^l_{m})/2,
\end{equation}
respectively. 
These currents are  governed by the AC-link voltage $V^r_m$, the DC-link voltage $V_{dc}$, and output and circulating arm voltage, which are defined as
\begin{equation}
     u^{\Delta}_{m}  =  (u^u_{m}- u^l_{m})/2, \quad \text{and} \quad u^{\Sigma}_{m}  = (u^u_{m}+ u^l_{m})/2,
\end{equation}
respectively. 
In the remainder, the superscripts ${(\cdot)}^{\Delta}$ and ${(\cdot)}^{\Sigma}$ denote variables associated with output currents (OC) and circulating currents (CC), respectively.

Considering the average MMC dynamics\footnotemark in the $dqz$ reference frame, as in \cite{Bergna2018_MMCModeling}, the application of Kirchhoff voltage law (KVL), yields the follow the differential equations: 
\begin{subequations}
    \label{eq:differential_equation_OCC_MMC}
    \begin{align}
        \label{eq:differential_equation_OCC_MMC_1}
        \frac{d}{dt} i^{\Delta}_d &= -\frac{R^{eq}_r}{L^{eq}_r} i^{\Delta}_d - \omega i^{\Delta}_q + \frac{1}{L^{eq}_r} u^{\Delta}_d - \frac{1}{L^{eq}_r} V^{r}_d,\\
        \label{eq:differential_equation_OCC_MMC_2}
        \frac{d}{dt} i^{\Delta}_q &= -\frac{R^{eq}_r}{L^{eq}_r} i^{\Delta}_q + \omega i^{\Delta}_d + \frac{1}{L^{eq}_r} u^{\Delta}_q - \frac{1}{L^{eq}_r} V^{r}_q,\\
          \label{eq:differential_equation_CCC_MMC_1}
       \frac{d}{dt} i^{\Sigma}_d &= -\frac{R_m}{L_m} i^{\Sigma}_d + 2 \omega i^{\Sigma}_q - \frac{1}{L_m} u^{\Sigma}_d,\\
        \label{eq:differential_equation_CCC_MMC_2}
        \frac{d}{dt} i^{\Sigma}_q &= -\frac{R_m}{L_m} i^{\Sigma}_q - 2 \omega i^{\Sigma}_d - \frac{1}{L_m} u^{\Sigma}_q,
    \end{align}
\end{subequations}  
where $L^{eq}_r = L_r+ \frac{L_m}{2},$ and $R^{eq}_r = R_r+ \frac{R_m}{2},$ and  $\omega = 2 \pi f$ is the angular velocity $f$ being the grid nominal frequency.
\footnotetext{Average dynamics assume that the insertion index (modulated control input) is a continuous signal and that the capacitor voltage is balanced \cite{Sharifabadi2016MMCBook}.}

\begin{remark}
    This paper focuses on modeling parameter uncertainties. Accordingly, MMC parameters are decomposed into nominal and uncertain components as follows: $f=f_0 + f_{\Delta}$, $L_m = L_{m0} +  L_{m\Delta}$, $R_m = R_{m0} + R_{m\Delta },$ $L_r = L_{r0} +  L_{r\Delta},$ and $R_r = R_{r0} + R_{r\Delta}$. In the remainder, the subscript ${(\cdot)}_{\cdot 0}$ and ${(\cdot)}_{\cdot \Delta}$ denote the nominal and uncertain value, respectively.
\end{remark}

\begin{assumption}[Uncertainty boundedness]
\label{assmp:uncertainty}
    All uncertain components are assumed to be unknown but bounded.    
\end{assumption}

% ============================================================================ %
\subsection{State-Space Realization}
% ============================================================================ %
The MMC current dynamics can be represented by uncertain state-space realizations (SSRs), which are affine combinations of nominal dynamics and uncertain dynamics, as in \cite{Shetgaonkar2023_MPCHVDC,Bergna2018_MMCModeling}.
We define a state vector $x^{\Delta}_{dq}  {:=} \begin{bmatrix} i^{\Delta}_d\; i^{\Delta}_q\; i^{\Sigma}_d\; i^{\Sigma}_q\end{bmatrix}^\top$, and a control input vector $u^{\Delta}_{dq}  :=  [u^{\Delta}_d{-}V^{r}_d\; u^{\Delta}_q{-}V^{r}_q\;  u^{\Sigma}_d\;  u^{\Sigma}_q]^\top$.
Based on~\eqref{eq:differential_equation_OCC_MMC},  a discrete-time uncertain SSRs yields
\begin{subequations}
\label{eq:discreteTimeSSMMC}
    \begin{align}
        {{x}_{dq}}^{+} &= (\textbf{A}_0+\textbf{A}_{\Delta}(\rho)) {x}_{dq} + (\textbf{B}_0+\textbf{B}_{\Delta}(\rho))u_{dq},\\
        y_{dq} &= \textbf{C} {x}_{dq},
    \end{align}
\end{subequations}
where $\textbf{A}_0 {:=} \operatorname{diag}(\textbf{A}^{\Delta}_0, \textbf{A}^{\Sigma}_0)$,  $\textbf{A}_{\Delta}(\rho) {:=} \operatorname{diag}(\textbf{A}^{\Delta}_{\Delta}, \textbf{A}^{\Sigma}_{\Delta})$,  $\textbf{B}_0:= \operatorname{diag}(\textbf{B}^{\Delta}_0, \textbf{B}^{\Sigma}_0)$ and $\textbf{B}_{\Delta}(\rho) {:=} \operatorname{diag}(\textbf{B}^{\Delta}_{\Delta}, \textbf{B}^{\Sigma}_{\Delta})$. The nominal dynamics are defined by $\textbf{A}^{\Delta}_0 {=} e^{\textbf{A}^{\Delta}_{c0} T_s}$, $\textbf{A}^{\Sigma}_0 {=} e^{\textbf{A}^{\Sigma}_{c0} T_s}$, $\textbf{B}^{\Delta}_0 {=} \int_0^{T_s}  e^{\textbf{A}^{\Delta}_{c} \tau} \textbf{B}^{\Delta}_{c0} d\tau$ and $\textbf{B}^{\Sigma}_0 {=} \int_0^{T_s}  e^{\textbf{A}^{\Sigma}_{c} \tau} \textbf{B}^{\Sigma}_{c0} d\tau$ with 
\begin{equation*}
\label{eq:continousTimeSSMMC_Nominal_Matrices}
    \begin{aligned}
        \textbf{A}^{\Delta}_{c0} &= \begin{bmatrix}
        -\frac{R_{eq0}}{L_{eq0}} & -\omega_{0} \\
            \omega_{0} & -\frac{R_{eq0}}{L_{eq0}}
        \end{bmatrix}, \, 
        \textbf{A}^{\Sigma}_{c0} = \begin{bmatrix}
        -\frac{R_{m0}}{L_{m0}} & 2\omega_0 \\
            -2\omega_0 & -\frac{R_{m0}}{L_{m0}}
        \end{bmatrix}, 
    \end{aligned}
\end{equation*}
$\textbf{B}^{\Delta}_{c0} = \operatorname{diag}(\frac{1}{L_{eq0}},\frac{1}{L_{eq0}})$, $\textbf{B}^{\Sigma}_{c0} = \operatorname{diag}(-\frac{1}{L_{m0}},-\frac{1}{L_{m0}})$ and $\textbf{C}=\textbf{I}_4$\footnote{$\textbf{I}_n$ denote the identity matrix of dimension $n$, $\operatorname{diag}(x_1,...,x_n)$ denote a diagonal matrix with $x_1$ to $x_n$ as diagonal elements. }.
The uncertain dynamics are defined by matrix functions
\begin{equation}
\label{eq:continousTimeSSMMC_Uncertaint_Matrices}
    \begin{aligned}
        \textbf{A}^{\Delta,\Sigma}_{\Delta}(\rho) &= \begin{bmatrix}
        \rho_{1} & \rho_{2} \\
            \rho_{3} & \rho_{4}
        \end{bmatrix}, \; \text{and} \;  \textbf{B}^{\Delta,\Sigma}_{\Delta}(\rho) = \begin{bmatrix}
            \rho_{5} & 0\\ 0 & \rho_{6} 
        \end{bmatrix}
    \end{aligned}
\end{equation}
with $\rho{:=}\left[\rho_1\; ...\; \rho_6\right]^{\top} {\in} \mathbb{P}{\subset}  \mathbb{R}^{6}$ being a vector with uncertain components.
Following Assumption~\ref{assmp:uncertainty}, the uncertain components are assumed to lie within a bounded polytopic convex set, denoted by $\mathbb{P}$. The parameter $\rho$ can capture variations due to component tolerances, as well as unmodeled dynamics arising from discretization effects and neglected nonlinearities.

\begin{remark}
    We abuse $\textbf{A}^{\Delta,\Sigma}_{\Delta}(\rho),  \textbf{B}^{\Delta,\Sigma}_{\Delta}(\rho)$ notation to declare that $\rho$ is the same in OC and CC.
    If $(\textbf{A}^{\Delta}_{\Delta}(\rho), \textbf{B}^{\Delta}_{\Delta}(\rho)) \neq \textbf{A}^{\Sigma}_{\Delta}(\rho), \textbf{B}^{\Sigma}_{\Delta}(\rho))$ is considered, the results hold too by increasing the amount of uncertain variables in $\rho$.
\end{remark}

% ============================================================================ %
% \subsection{Augmented State-Space Realization with Integral Action}
% ============================================================================ %
To enable offset-free tracking of piecewise-constant setpoints, an augmented SSR is used.
Let's define a new augmented state vector, i.e., $\vec{x}_{dq}:= \begin{bmatrix} \delta {x}_{dq} & y_{dq} \end{bmatrix}^{\top}$; then, the augmented MMC SSR yields  
\begin{subequations}
\label{eq:discreteTAugementedTimeSSMMC}
    \begin{align}
         {{\vec{x}}^{+}_{dq}} &= \vec{\textbf{A}}(\rho )\vec{x}_{dq} + \vec{\textbf{B}}(\rho ) \delta u_{dq},
        % y_{dq} &= \textbf{C} \vec{x}_{dq},
    \end{align}
\end{subequations}
with $\vec{\textbf{A}} := \vec{\textbf{A}}_0 +\vec{\textbf{A}}_{\Delta}(\rho )$ and  $\vec{\textbf{B}} := \vec{\textbf{B}}_0 +\vec{\textbf{B}}_{\Delta}(\rho )$,
where
{\small
\begin{equation*}
\label{eq:discreteTAugementedTimeSSMMC_Matrices}
\vec{\textbf{A}}_0 {=} \begin{bmatrix}
        \textbf{A}_0 & \boldsymbol{0}_2\\
        \textbf{C} \textbf{A}_0  & \textbf{I}_2
    \end{bmatrix}, 
\vec{\textbf{A}}_{\Delta} {=} \begin{bmatrix}
        \textbf{A}_{\Delta} & \boldsymbol{0}_2\\
        \textbf{C} \textbf{A}_{\Delta} & \boldsymbol{0}_2
    \end{bmatrix} , 
    \vec{\textbf{B}}_0{=}\begin{bmatrix}
        \textbf{B}_0\\
       \textbf{C} \textbf{B}_0
    \end{bmatrix}, 
\vec{\textbf{B}}_{\Delta}{=} \begin{bmatrix}
        \textbf{B}_{\Delta}\\
       \textbf{C} \textbf{B}_{\Delta}
    \end{bmatrix}
\end{equation*}
}
$\delta {x}_{dq}(k) {:=} {x}_{dq}(k) - {x}_{dq}(k{-}1)$ and $\delta u_{dq}(k) {:=} u_{dq}(k) - u_{dq}(k{-}1)$.
In the reminder, the operator $\delta$ denotes the discrete-time derivative, and the vector notation $\vec{\cdot}$ refers to the augmented SSR components, such as state vectors and matrices.

% ============================================================================ %
\section{MMC Control System}
\label{sec:Control_Problem}
% ============================================================================ %

\begin{figure}[htp]
    \centering
    \includegraphics[width=\linewidth]{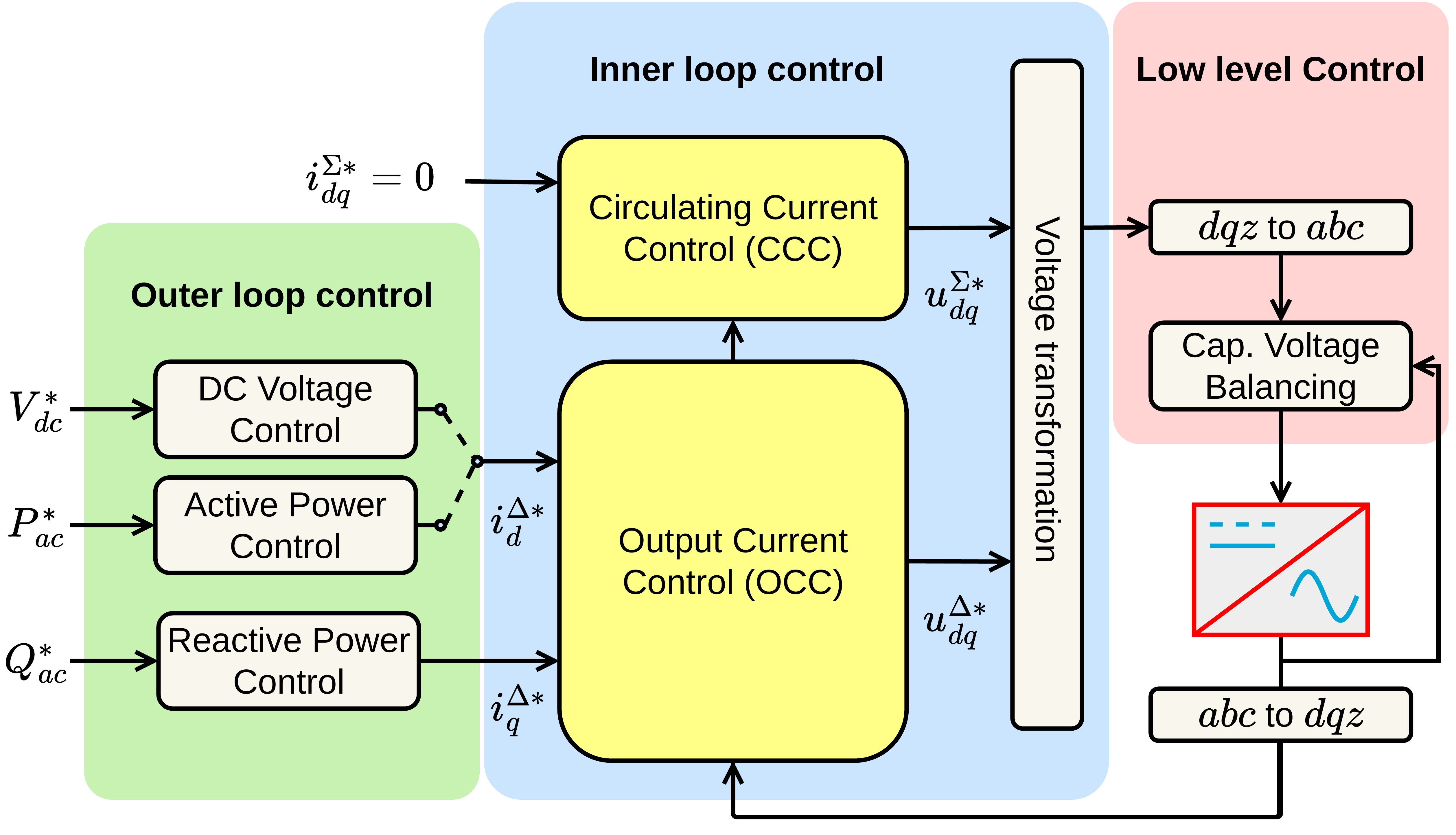}
    \caption{Hierarchical decoupled control architecture of MMC currents.}
    \label{fig:Hierarchical_Control_Diagram}
\end{figure}

The control architecture of an MMC is commonly organized in a hierarchical structure comprising outer, inner, and low-level control loops, as shown in \cref{fig:Hierarchical_Control_Diagram}. 
The outer loop receives $(P^*_{ac}, Q^*_{ac})$ or $V^*_{dc}$ references from the transmission system operator (TSO) or supervisory control, depending on the operating mode, and generates piecewise constant current references $i^{\Delta *}_{dq}$ based on the PQ power equations as in \cite{IEEE_STD_2010}. 
These references typically are required to be settled within $0.2$--$1.5\,\mathrm{s}$~\cite{NSWPH_MTHVDC_2023}. 
The circulating current governs the MMC energy balance; here, $i^{\Sigma *}_{dq}=0$ is assumed to ensure proper energy balancing.
Finally, the inner-loop controllers generate voltage references for the low-level control. 

A common design approach for the inner control loop employs cascade PI controllers in the synchronous $dq$ reference frame, where constraint satisfaction is typically enforced through saturation blocks and current limiters, see e.g. \cite{Tavakoli2022_RobustHVDC}.
As an alternative, we propose replacing the PI-based structure with an MPC  augmented by a feedforward term.
The feedforward component is designed to compensate for steady-state behavior and reduce the MPC control effort, and is given by
\[
K^{\text{ff}} = \mathbf{B}_0^{-1}(I - \mathbf{A}_0)
\begin{bmatrix}
i^{\Delta *}_{dq} \\
i^{\Sigma *}_{dq}
\end{bmatrix}.
\]
This formulation ensures offset-free tracking under nominal conditions and enhances the dynamic response by pre-shaping the control input according to the reference trajectories.

\begin{figure}[htp]
    \centering
    \includegraphics[width=0.85\linewidth]{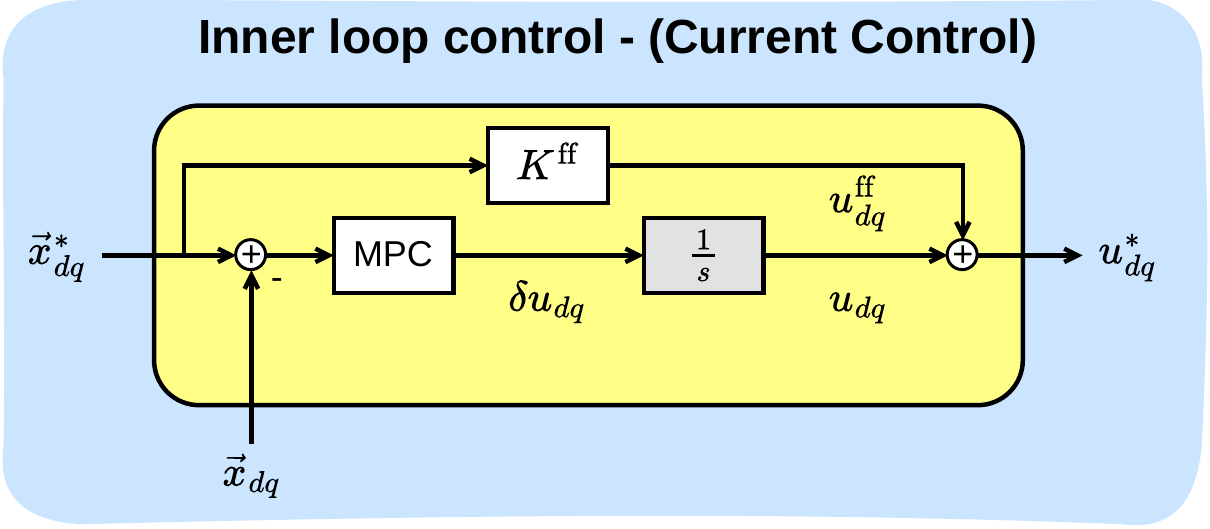}
    \caption{MPC-based control architecture for the inner loop. }
    \label{fig:Diagram_Current_Control_Comp}
\end{figure}

\section{Robust Model Predictive Control for MMC-based MMC}
\label{sec:RMPC_Design}

\begin{figure}[htp]
    \centering
    \includegraphics[width=\linewidth]{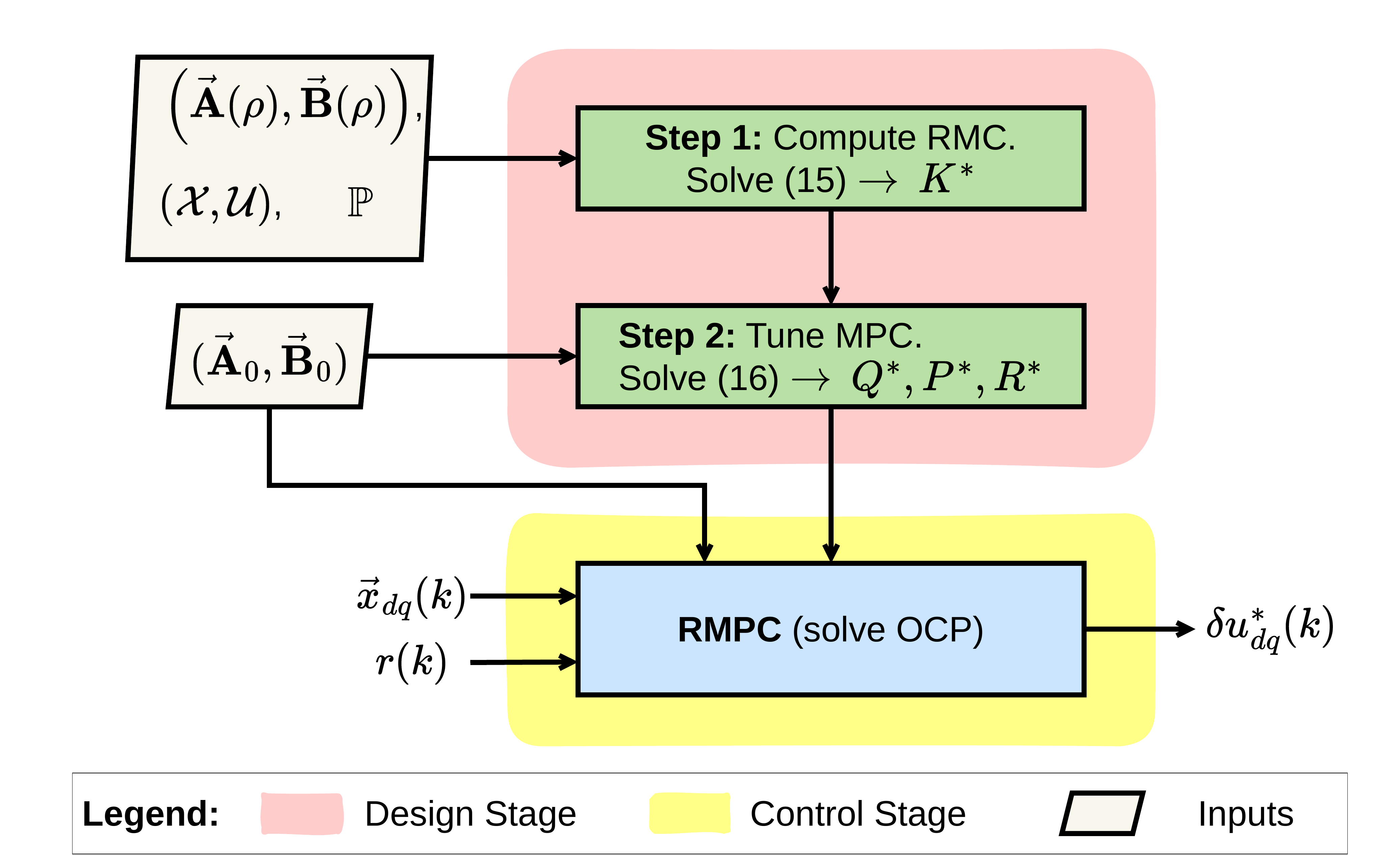}
    \caption{MPC design based on robust matching controller.}
    \label{fig:Control_Diagram_RMPC_ver1}
\end{figure}

The proposed MPC robust tuning method leverages a matching-controller tuning algorithm from \cite{ReyesDreke_DBMC_2023} in combination with a robust matching controller (RMC) of our design. 
Accordingly, we explain in this section: (\emph{i}) general MPC formulation, (\emph{ii}) RMC design, and (\emph{iii}) automatic tuning via matching with vertex relaxation.
\cref{fig:Control_Diagram_RMPC_ver1} illustrates a diagram summarizing our design methodology. 

\subsection{MPC Formulation}
Classic MPC relies on prediction model that considers only rated values, i.e., a nominal prediction model. 
At time $k$, for a prediction horizon $N$, discrete-time step ahead predictions are described by:
\begin{equation}
    \vec{x}_{i+1 \mid k}= \vec{\textbf{A}}_0  \vec{x}_{i \mid k}+   \vec{\textbf{B}}_0 \delta u_{i \mid k}, \quad \forall i \in \{0,{...}, N\},
\end{equation}
where $ \vec{x}_{0 \mid k} =  \vec{x}(k)$.
The predicted dynamics at time $k$ are stacked into vectors, such that $X_k {=}[ \vec{x}^{\top}_{0 \mid k} \ldots  \vec{x}^{\top}_{N \mid k}]^{\top}$ and $U_k{=}[\delta u^{\top}_{0 \mid k} \ldots \delta u^{\top}_{N \mid k}]^{\top}$ are the state and input stacked vectors, respectively.
Similarly, the references signals $r$ are staked into vector $R_k = [ r^{\top}(k) \ldots  r(k+N))]^{\top}$, where, for all $k$, the references vectors have a fixed structure, i.e.,

\begin{equation}
    r = \begin{bmatrix} 0 &0 & {i}^{\Delta *}_{d} & {i}^{\Delta *}_{q} &0 &0 & {i}^{\Sigma *}_{d} & {i}^{\Sigma *}_{q}\end{bmatrix}^\top.
\end{equation}  
Classic MPC solves at time $k$, the following finite-horizon optimal control problem (OCP) 
\begin{subequations}
\label{eq:MPC_Problem}
\begin{align}
\label{eq:MPC_Problem_value_function}
\min _{U_k}\; &\left(X_k-R_k\right)^{\top}\Omega\left(X_k-R_k\right) + U_k^{\top}\Psi U_k, \\
\label{eq:MPC_Problem_equality_constraints_model}
\text { s.t. }   X_{k+1}&=\Phi  x_{0 \mid k}+  \Gamma U_{k},  \\
\label{eq:MPC_Problem_inequality}
 x_{i \mid k}&\in\mathbb{X}, \;u_{i \mid k}\in\mathbb{U}, \;\delta u_{i \mid k}\in\mathbb{U}_{\delta},\; \forall i {\in} \{0,...,N\},\\
\label{eq:MPC_Problem_inequality_constraints_terminal_set}
  x_{N \mid k} &\in \mathcal{X}_T \subseteq \mathbb{X},
\end{align}
\end{subequations}
where $X_{k+1} {=}[ \vec{x}^{\top}_{1 \mid k} \ldots  \vec{x}^{\top}_{N+1 \mid k}]^{\top}$,  $\Omega = \operatorname{diag}(Q,...,Q,P) \in \mathbb{R}^{8N\times 8N}$ and $\Psi = \operatorname{diag}(R,...,R)\in \mathbb{R}^{4N\times 4N}$.
We assume that  {weighting matrices} $Q, P \in \mathbb{R}^{8 \times 8}$, $R \in \mathbb{R}^{4 \times 4}$ satisfy the conditions $Q=Q^{\top} \succeq 0, P=P^{\top} \succeq 0$, $R=R^{\top} \succ0$. 
Equation \eqref{eq:MPC_Problem_equality_constraints_model} calculates a step-ahead prediction of the stacked state $X_{k+1}$ using the matrices $\Gamma$ and $\Phi$ defined as
\begin{equation*}
\label{eq:prediction_matrices}
{\Gamma}{=}\left[\begin{array}{cccc}
 \vec{\textbf{B}}_0 & 0 & \ldots & 0 \\
 \vec{\textbf{A}}_0  \vec{\textbf{B}}_0 &  \vec{\textbf{B}}_0 & \ldots & 0 \\
\vdots & \vdots & \ddots & \vdots \\
 \vec{\textbf{A}}_0^{N-1}  \vec{\textbf{B}}_0 & \vec{\textbf{A}}_0^{N-2}  \vec{\textbf{B}}_0 & \ldots &  \vec{\textbf{A}}_0^{N}\vec{\textbf{B}}_0
\end{array}\right], \, {\Phi}{=}\left[\begin{array}{c}
 \vec{\textbf{A}}_0 \\
\vec{\textbf{A}}_0^2 \\
\vdots \\
 \vec{\textbf{A}}_0^N
\end{array}\right]
\end{equation*} 
The state and control input, control input rate constraints formulated in \eqref{eq:MPC_Problem_inequality}, are defined by the sets $\mathbb{X} =\{\vec{x}\in \mathbb{R}^8: \vec{x}_{\text{min}} \leq \vec{x} \leq \vec{x}_{\text{max}}\},$  $\mathbb{U} =\{u\in \mathbb{R}^4: u_{\text{min}} \leq u\leq u_{\text{max}}\},$ and $\mathbb{U}_{\delta} =\{\delta u\in \mathbb{R}^4: \delta u_{\text{min}} \leq \delta u\leq \delta u_{\text{max}}\},$ respectively. Condition \eqref{eq:MPC_Problem_inequality_constraints_terminal_set} describes the terminal constraints where $\mathcal{X}_T \subseteq \mathbb{X}$ is a terminal set that should satisfy the standard conditions for stability and recursive feasibility presented in \cite{Rawlings2008model}. 

If no constraints are active, the OCP yields an optimal unconstrained solution (OUS), i.e.,
\begin{equation}
\label{eq:MPC_Unconstrainted_Solution_Uopt}
U^*_{k}=\left[\begin{array}{ccc}
\delta u^{*\top }_{0 \mid k} &
\ldots &
\delta u^{*\top }_{N-1 \mid k} 
\end{array}\right]^{\top}=-G^{-1} F(x(k))
\end{equation} 
with $G =2\left(\Psi+\Gamma^{\top} \Omega \Gamma\right)$ and $F(x(k)) = 2 (\Gamma^{\top} \Omega \left( \Phi x(k) - {R}_k\right))$. To avoid confusion, we denote the OUS by $\delta\hat{u}_{0 \mid k}$

\begin{remark}
    For finite prediction horizons, MPC closed-loop stability is not guaranteed for arbitrary choices of $(Q, R, P)$. In fast applications, the prediction horizon $N$ must remain small; therefore, stability critically depends on the computation of $P$, i.e., the terminal cost and the associated terminal set $\mathcal{X}_T$.
\end{remark}

\subsection{Robust Matching Controller }
% =====================================================================
The proposed RMC is a static state-feedback controller designed to be robust against model uncertainties within a predefined uncertainty set. Unlike existing approaches, its design explicitly incorporates soft constraints, aiming to enlarge the region of attraction in which the resulting stabilizing control input trajectories remain constraint-admissible.

Let us define the uncertain set $\mathbb{P}$ as a convex hull set such that
\begin{equation}
    \label{eq:polytope_defintion}
    \mathbb{P} =  \text{conv} \{ v^1, v^2, \ldots, v^p\},
\end{equation}
where $p \in \mathbb{N}_+$, $v^j$  are the $j$th vertex of the polytope $\mathbb{P}$ and $\text{conv} \{ \cdot\}$ denotes the operation of taking the convex hull of the argument. 

Then, we define a set of soft constraints for the tracking error to design the RMC such that
\begin{subequations}
    \label{eq:constraintDefinitionLemma}
        \begin{align}
            \mathcal{X} &:= \{e_x\in\mathbb{R}^{8}: g_{t_x}^\top e_x \leq 1, \forall t_x \in \{1,{...},s\}\},\\
            \mathcal{U} &:= \{ e_u\in\mathbb{R}^{8}: h_{t_u}^\top  e_u \leq 1, \forall t_u \in \{1,{...},l\}\},
        \end{align}
\end{subequations}
such that
\begin{equation}
    \begin{aligned}
    &e_x +  r \in {\mathbb{X}}, \; \forall e_x \in \mathcal{X}, \; \text{and} \; e_u +  u^{s} \in {\mathbb{U}}, \; \forall e_u \in \mathcal{U}
    \end{aligned}
\end{equation}
where $e_x(k):= \vec{x}-r$ and $e_x(k) \in \mathcal{X} \implies \vec{x} \in \mathbb{X}$, $u^s= 0_{2\times 1}$, $e_u(k):= \delta u-u^s$ and $e_u(k) \in \mathcal{U} \implies u \in \mathbb{U}$.

To find a RMC, we propose the following LMIs:
\begin{subequations}
\label{eq:constrained_synthesis}
    \begin{align}
    \label{eq:constrained_synthesis_Lyapunov}
    \begin{aligned}
        &\forall j \in \{1, ..., p\},\\
        &\begin{bmatrix}
        Z & (\vec{\textbf{A}}(v^j)Z+\vec{\textbf{B}}(v^j)Y)^{\top} \\
        (\vec{\textbf{A}}(v^j)Z+\vec{\textbf{B}}(v^j)Y) & Z
        \end{bmatrix} \succ 0, 
    \end{aligned}&  \\
    \label{eq:constrained_synthesis_constraint_x}
        \forall t_x \in\left\{1, \ldots, s\right\}, \quad \begin{bmatrix}
            Z & \left(Z g_{t_x}\right)^{\top} \\
            \left(Z g_{t_x}\right) & 1
        \end{bmatrix}  \succcurlyeq 0,& \\
        \label{eq:constrained_synthesis_constraint_u}
        \forall t_u \in\left\{1, \ldots, l\right\}, \quad \begin{bmatrix}
             Z & \left(Y h_{t_u}\right)^{\top} \\
            \left(Y h_{t_u}\right) & 1
        \end{bmatrix}\succcurlyeq 0,& 
    \end{align}        
\end{subequations}
where  $K =YZ^{-1}$ defines the RMC, $g_{t_x}$ and $h_{t_u}$ are normalized vectors corresponding to the error constraints from $\mathcal{X}$ and $\mathcal{U}$, respectively, $p$ is number of vertices of $\mathbb{P}$, $v^j$ are the vertices of $\mathbb{P}$.

\begin{remark}
    Equation \eqref{eq:constrained_synthesis_Lyapunov} expresses the discrete-time Lyapunov stability condition as LMIs. 
    Their feasibility guarantees the existence of a quadratic Lyapunov function 
    $V(\vec{x})=\vec{x}^\top P \vec{x}$, $P \succ 0$, and a static RMC gain $K$ that ensures 
    robust exponential stability for all admissible uncertainty realizations within 
    the polytopic set $\mathbb{P}$. For affine, bounded uncertainties, robust stability over $\mathbb{P}$ can be verified by checking the LMIs at the vertices of the uncertainty polytope, leveraging convexity arguments and the Schur complement (see, e.g., \cite{Boyd1994_LMIBook,Blanchini1999_SetInvariance}).
    
    Equations~\ref{eq:constrained_synthesis_constraint_x}-~\ref{eq:constrained_synthesis_constraint_u} enforce soft state and input constraints by embedding a robust positively invariant ellipsoid inside the admissible polytopic sets. This is achieved using standard ellipsoidal inclusion conditions, ensuring $\vec{x}^\top P \vec{x} \le \alpha \Rightarrow \vec{x} \in \mathbb{X}$ and $K\vec{x} \in \mathbb{U}_{\delta}$ (see \cite{Blanchini1999_SetInvariance,Kolmanovsky1998_MPC}).
\end{remark}

\begin{remark}
    The computed RMC, denoted by $K$, admits a domain of attraction (DOA) defined as
\[\mathbb{S} := \{ \vec{x} \in \mathbb{R}^8 : \vec{x}^\top P \vec{x} \leq 1,\;  K\vec{x} \in \mathcal{U} \} \subseteq \mathcal{X}.\]
This DOA is a hyperellipsoidal set that is also positively invariant and constraint-admissible. Therefore, once the system state enters $\mathbb{S}$, convergence to the steady state is guaranteed to be robust against perturbations within $\mathcal{X}$, while ensuring that the control input remains bounded within the admissible set $\mathcal{U}$. Based on the DOA, we can infer the transient response.
\end{remark}

\begin{remark}
    To maximize the dimension of the DOA, the following optimization problem is  formulated
     \begin{equation}     \label{eq:minimization_problem_LMI_constrait}
     \begin{aligned}
         \min_{Z,Y} \; &-\det(Z)^{1/n} \\
         \text{s.t.}&  \; \eqref{eq:constrained_synthesis}.
     \end{aligned}         
     \end{equation}
     where $n$ is the dimension of $Z$. Note that $-\det(Z)^{1/n}$ is convex operation that allow us to maximize the determinant of $Z$ and consequently, minimize the eigenvalues of $P$. The result of \eqref{eq:minimization_problem_LMI_constrait} is the optimal RMC denoted as $K^*$.
     % Readers are referred to \cite{ReyesDreke_DBMC_2023} to consult examples of how to solve this optimization problem using MATLAB/MOSEK\cite{mosek}. 
\end{remark}

\subsection{MPC Tuning via Controller Matching}
% ----------------------------------------------------------------------------------------------------------- %
MPC tuning controller matching are set of methods to computed optimal weighting matrices $Q,R,P$, such that the unconstrained response of the MPC mimics the properties of pre-computed controller, e.g., robustness, transient response or stability. 
In this case, the pre-computed controller is the RMC. 
Following the methodology presented in  \cite{ReyesDreke_DBMC_2023}, given a RMC $K$ computed as in \eqref{eq:constrained_synthesis}, MPC tuning via controller matching is solved by
\begin{subequations}
    \label{eq:optimization_matching_controller_with_relaxation}
        \begin{align}
          \label{eq:cost_function_vertex_relaxation}
            J&=\min _{P,Q,R,\varepsilon} \varepsilon
        \end{align}
        \begin{multline}
          % \label{eq:lmi_norm_vertex_relaxation}
          \text {s.t.}  \left\|\left(\left({\Psi}+{\Gamma}^{\top} \Omega {\Gamma}\right) {\mathcal{K}}+{\Gamma}^{\top} \Omega {\Phi}\right) \cdot \xi^j\right\| \leq \varepsilon, \\ \forall j=\{1, \ldots, p\},
          \end{multline}
          \begin{align}
            % \label{eq:stability_guarantee_vertex_relaxation}            
            &(\vec{\textbf{A}}_0+\vec{\textbf{B}}_0 K)^{\top} P(\vec{\textbf{A}}_0+\vec{\textbf{B}}_0 K)+K^{\top} R K+Q-P {\preccurlyeq} 0, \\
            % \label{eq:matrix_signs_vertex_relaxation}
            &Q \succ 0, \; P \succ 0 , \; R \succeq \sigma I, \; \varepsilon \geq 0,\; \sigma>0.
        \end{align}
    \end{subequations} 
where $\xi^j$ are the vertices of the constraint set $\mathbb{X}$ and 
\begin{equation}
    \label{eq:caligraph_K}
    \mathcal{K}=\left[\begin{array}{c}
    K \\
    K(\vec{\textbf{A}}_0+\vec{\textbf{B}}_0 K) \\
    \vdots \\
    K(\vec{\textbf{A}}_0+\vec{\textbf{B}}_0 K)^{N-1}
    \end{array}\right].
\end{equation}
The result of \eqref{eq:optimization_matching_controller_with_relaxation} are the optimal $Q,R$ and $P$ denoted as $Q^*, R^*$ and $P^*$.
\begin{remark}
    The optimization problem in \eqref{eq:optimization_matching_controller_with_relaxation} can be solved using MATLAB/MOSEK~\cite{mosek}. Since the problem is convex and linear, global optimality is guaranteed. However, for high-dimensional systems, the number of constraints may grow exponentially, resulting in increased computational burden. In such cases, the approach presented in \cite{Cariano2009_MPC} is recommended. 
    Problems~\eqref{eq:minimization_problem_LMI_constrait} and~\eqref{eq:optimization_matching_controller_with_relaxation} were solved on a 13th Gen Intel(R) Core(TM) i7-1365U processor with 16\,GB of RAM. The average computation times were 1.93\,s and 0.139\,s, respectively.
    % Readers are referred to \cite{ReyesDreke_DBMC_2023} for illustrative examples and implementation details using MATLAB/MOSEK.
\end{remark}

% ===================================================== %
\section{Case of Study}
\begin{figure*}[!htp]
    \centering
    \includegraphics[width=\linewidth]{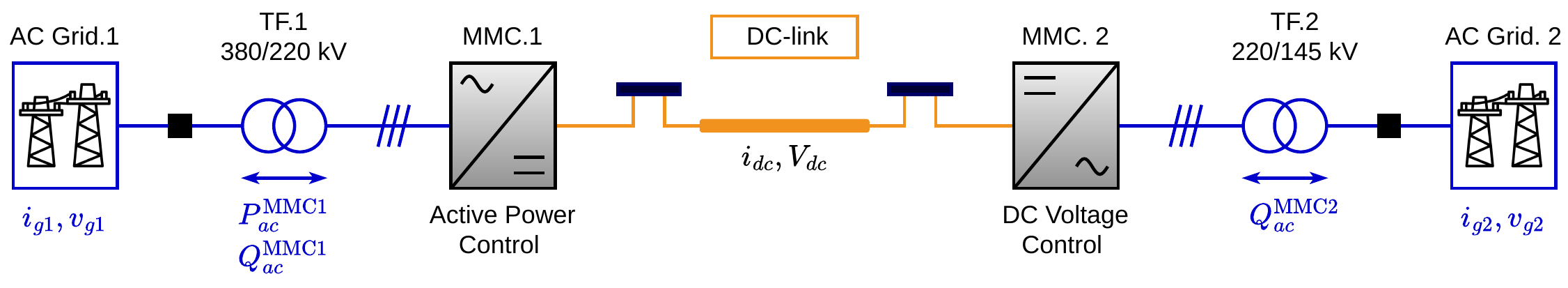}
    \caption{Simplified schematic of an MMC-based HVDC network with a point-to-point mono-polar connection.}
    \label{fig:MTDC_Diagrram_Two_Terminals}
\end{figure*}

% ===================================================== %
\label{sec:Study_Case}
The effectiveness of the proposed method is evaluated on a point-to-point monopolar HVDC network, as shown in \cref{fig:MTDC_Diagrram_Two_Terminals}. 
The HVDC system is based on the CIGRE benchmark model~\cite{Vrana2013_Cigre}, and the rated power, along with the nominal parameters of the components, are reported in Table~\ref{tab:Tennet_MTDC_2GW_parameters}. 
In this setup, power can be transmitted bidirectionally from Grid 1 to Grid 2. Converter MMC.1 operates in active power control mode, while MMC.2 regulates the DC voltage.
The outer controllers are kept as PI controllers with $K_p = 8$ and $T_i = 0.00367$.
The low-level controllers are outside of the scope, and the original from the benchmark is used.

\begin{table}[htp]
    \centering
    \caption{HVDC rated parameters \cite{Vrana2013_Cigre}.}
    \begin{tabular}{|l|l|l|l|}
    \hline Parameter & Value & Parameter & Value \\
    \hline Rated power & 800 MVA & Rated frequency & 50 Hz \\
    \hline MMCs AC voltage & 220 kV & MMCs DC voltage & 400 kV \\
    \hline AC grid 1 voltage & 380 kV & AC Grid 2 voltage & 145 kV \\
    \hline Number of SMs & 150 & SM capacitance & 10 mF \\
    \hline SM switching voltage & 3.5 kV & SM switching current & 1 kA \\
    \hline Arm resistance & $0.15 \Omega$ & Arm inductance & 29 mH \\
    \hline TFs resistance & 0.006 pu & TFs inductance & 0.18 pu \\
    \hline
    \end{tabular}
    \label{tab:Tennet_MTDC_2GW_parameters}
\end{table}
% We consider structured and unstructured uncertainties. 
Related to the structured uncertainties, the passive component values are considered to be between their 10\% to 300\% of their nominal  value, such that $L_{m\Delta} = (\alpha-1)L_{m0}$, $L_{r\Delta} = (\alpha-1)L_{r0}$, $R_{m\Delta} = (\alpha-1)R_{m0}$, and $R_{r\Delta} = (\alpha-1)R_{r0}$ with $\alpha \in [-0.9, 2]$. Similarly, the fundamental frequency is expected to vary with  $ f_{\Delta} \in [0, 0.1\% f_{0}]$.
The structured uncertainties can be bounded by an uncertainty set $\mathbb{P}$ constrained as
\begin{equation*}
\label{eq:scheduling_set}
    \begin{array}{lll}
         \rho_1 \in [-\frac{4}{10^3},\frac{4}{10^3}], &\rho_2 \in [-\frac{6}{10^3},\frac{6}{10^3}], &\rho_3 \in [-\frac{6}{10^3},\frac{6}{10^3}], \\ \rho_4 \in [-\frac{4}{10^3},\frac{4}{10^3}],
         &\rho_5 \in [-\frac{1}{10^{3}},\frac{1}{10^{3}}], &\rho_6 \in [-\frac{1}{10^{3}},\frac{1}{10^{3}}].
    \end{array}
\end{equation*}

From Table~\ref{tab:Tennet_MTDC_2GW_parameters}, we derive a per-unit model of the MMC dynamics. 
The RMPC design for the inner loop considers a prediction horizon $N=5$ and hard constraints on the first prediction step, such that $\left\|u_{k+1\mid k}\right\|_{\infty}\leq 1.1$ and $\left\|\delta u_{k+1\mid k}\right\|_{\infty}\leq 0.25$. 
The RMPC is implemented in RSCAD using a CBuilder block that executes the Hildreth algorithm in C to solve the OCP.

The RMC design considers the following soft constraints:
\begin{equation}
\label{eq:softConstraintDefinition}
    \begin{aligned}
        \mathcal{X}&:= \left\{\vec{x}^{}_{dq}\in\mathbb{R}^8: -5\textbf{1}_8 \leq \vec{x}^{}_{dq} \leq 5\textbf{1}_8 \right\}, \\ 
        \mathcal{U}&:= \left\{\delta{u}^{}_{dq}\in\mathbb{R}^4: -0.01\textbf{1}_4 \leq \delta{u}^{}_{dq} \leq 0.01\textbf{1}_4 \right\} ,
    \end{aligned}
\end{equation}
where $\textbf{1}_n$ defines a column vector fill with $n$ ones and $\leq$ operator defines element-wise comparison. 

Solving \eqref{eq:minimization_problem_LMI_constrait} and \eqref{eq:optimization_matching_controller_with_relaxation}, the RMC and optimal weighting matrices $(Q^*,R^*,P^*)$ presented Appendix~\ref{sec_appx:Gains} were obtained.

For a fair comparison, an LQR-based tuning is employed with $Q = 5\textbf{I}_8$ and $R = 20\textbf{I}_2$, yielding the terminal cost and unconstrained control law $(P_{LQR},K_{LQR})$ presented in the Appendix~\ref{sec_appx:Gains}. The matrices $Q$ and $R$ were selected empirically through trial-and-error, and have shown excellent nominal performance in previous works. 

\section{Control Design Verification}
\label{sec:Simulation_results}

% =====================================================================
% \textcolor{red}{Robust stability of standalone MMC is analyzed through eigenvalue analysis.}
% =====================================================================
The control design verification is performed in a closed loop with the derived uncertain per-unit model. 
Due to the nonlinear nature of MPC, classical stability assessment techniques are not directly applicable. 
However, MPC unconstrained responses can still be analyzed, since in both cases the OUS recovers the corresponding LQR and RMC responses, such that $\delta \hat{u}^{\text{LQR}}_{0\mid k} =(\vec{\textbf{A}}+\vec{\textbf{B}}K_{\text{LQR}})\vec{x}(k)$ and $\delta \hat{u}^{\text{RMPC}}_{0\mid k} =(\vec{\textbf{A}}+\vec{\textbf{B}}K^{*})\vec{x}(k)$, respectively.

\begin{figure}[htbp]
    \centering
    \includegraphics[width=\linewidth]{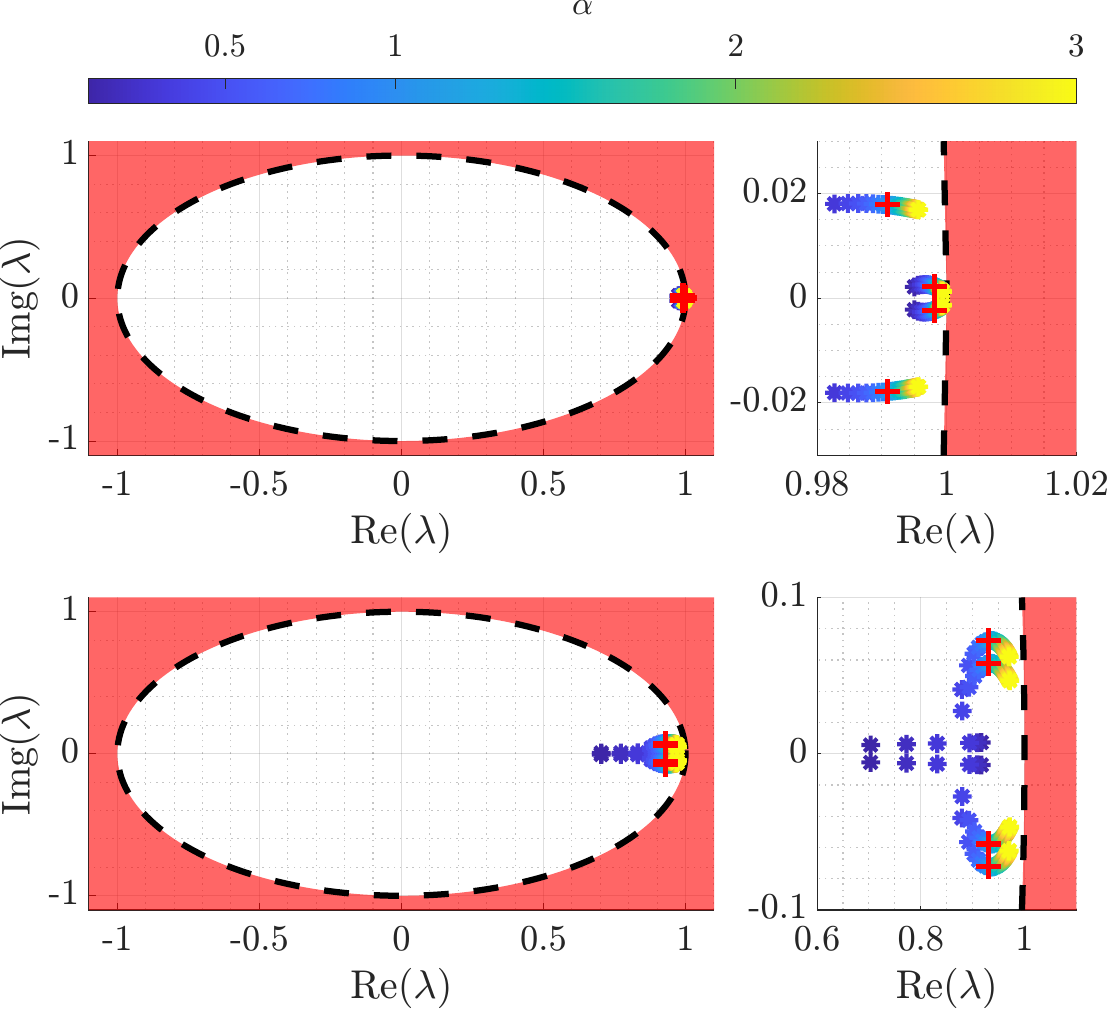}
    \caption{Closed-loop eigenvalues of the OC system under RMPC (top) and (bottom) LQR-MPC for different \( L_{r} = \alpha L_{r0} \) and \(  R_{r} = \alpha R_{r0} \). (Right) Zoomed view around the nominal eigenvalues, indicated by the red ‘\textcolor{red}{+}’ marker.}
    \label{fig:EignUncertOC_RMPC}
\end{figure}

Closed-loop stability is evaluated by analyzing the system eigenvalues $\lambda$ under uncertain parameters for different $\alpha$, $L_r$ and $R_r$, as shown in \cref{fig:EignUncertOC_RMPC}. 
Both unconstrained RMPC and LQR-MPC guarantee stability, as the eigenvalues remain within the unit circle.
% of $\vec{\textbf{A}}+\vec{\textbf{B}}K^{*}$ and $\vec{\textbf{A}}+\vec{\textbf{B}}K_{\text{LQR}}$  
The nominal eigenvalues are located in similar regions of the complex plane, with $\lambda^{\text{nom}}_{\text{RMPC}} = \{0.99\pm0.0178j,\; 0.998\pm0.002j\}$ and $\lambda^{\text{nom}}_{\text{LQR}} = \{0.93\pm0.072j,\; 0.93\pm0.058j\}$. Hence, similar closed-loop responses are expected under nominal conditions.
The RMPC eigenvalues remain closer to their nominal locations under uncertainty, indicating a response that is more consistent with the nominal case than LQR. 
Similar observations were obtained for all eigenvalues (both OC and CC) under simultaneous variation of all parameters within $\mathbb{P}$; however, these results are omitted due to space constraints.

\begin{figure}[htbp]
    \centering
    \includegraphics[width=\linewidth]{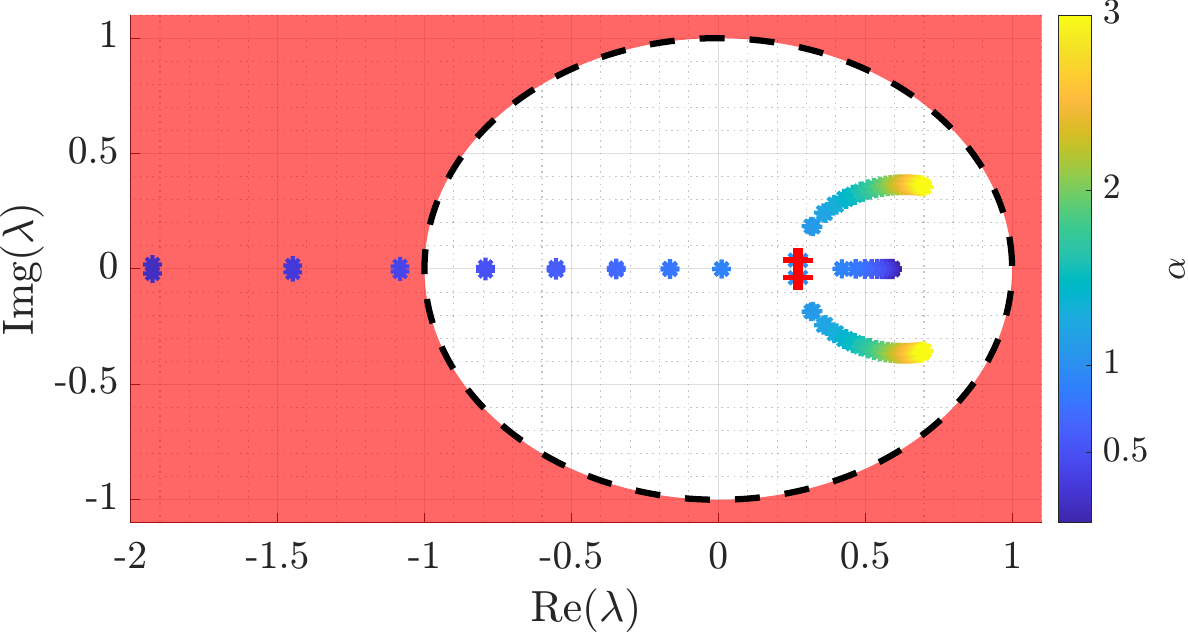}
    \caption{Closed-loop eigenvalues of the OC system under alternative LQR-MPC with $Q_2,R_2,P_2$ for different \( L_{r} = \alpha L_{r0} \) and \(  R_{r} = \alpha R_{r0} \) and nominal value indicated by the red ‘\textcolor{red}{+}’ marker.}
    \label{fig:EignUncertOC_LQR}
\end{figure}

To highlight the sensitivity of LQR-MPC robustness to the choice of $Q$ and $R$, we analyze the closed-loop eigenvalues using an alternative tuning $Q_2 {=} 10 \times \mathbf{I}_8$ and $R_2 {=} 10^{-2}\times\mathbf{I}_4$, resulting in $K^{2}_{\text{LQR}}$ (see Appendix~\ref{sec_appx:Gains}), which is expected to improve transient response. \cref{fig:EignUncertOC_LQR} shows eigenvalues of $\vec{\textbf{A}}+\vec{\textbf{B}}K^{2}_{\text{LQR}}$ under the uncertainty variation. 
Although the nominal eigenvalues are closer to the origin, indicating faster convergence, stability is not maintained for all uncertainty instances. 
In particular, simultaneous reductions in $L_r$ and $R_r$ can lead to instability. 
This behavior is attributed to the LQR controller's high gain. 
Here, reducing $L_r$ increases the effective system gain, potentially violating gain margins.

\begin{figure}[htbp]
    \centering
    \includegraphics[width=\linewidth]{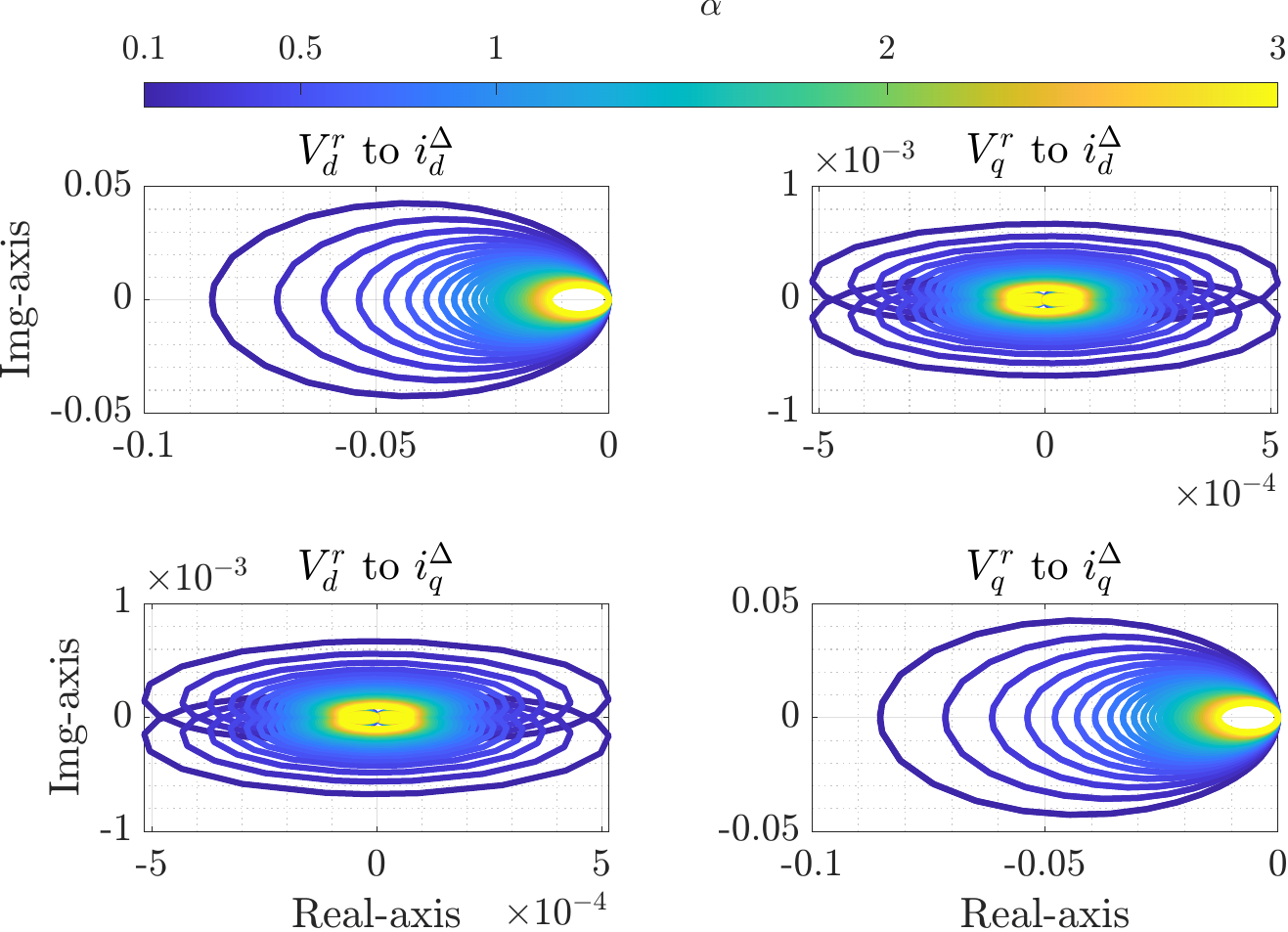}
    \caption{MIMO Nyquist plot of the OC system under RMPC for varying uncertainty levels in \( L_{r\Delta} \) and \( R_{mr\Delta} \). }
    \label{fig:NyquistUncertOC_RMPC_Full}
\end{figure}

% % =====================================================================
% \textcolor{red}{Robust stability of connect MMC is analyzed through Nyquist analysis.}
% % =====================================================================
To assess the robust stability of the RMPC and its interaction with the grid, the Nyquist plot of the MMC closed-loop dynamics is evaluated, considering the grid voltage as the input, as shown in \cref{fig:NyquistUncertOC_RMPC_Full}. 
The results indicate that, for all uncertainties within the set $\mathbb{P}$, the phase margin is effectively infinite, while the gain margin is significantly greater than 6 dB, which is commonly regarded as a minimum robustness requirement. 
These findings confirm the strong robustness properties of the proposed controller.

%%%%%%%%%%%%%%%%%%%%%%%%%%%%%%%%%%%%%%%%%%%%%%%%%
\section{RTDS simulation results}
The effectiveness of the controller is evaluated through electromagnetic transient (EMT) simulations in RTDS. 
At this stage, the designed controller must handle unmodeled phenomena in \eqref{eq:discreteTAugementedTimeSSMMC}, including disturbances, switching-induced nonlinear dynamics, and interactions with other control loops and filters. The RTDS simulations are executed on the RTDS NovaCor 2.0 rack, which is capable of running the RMPC CBuilder code within the sampling time \(T_s\).
Performance is assessed using time-domain step responses, frequency-domain analysis of the admittance matrix Bode plots, and harmonic analysis to evaluate power quality.
% This combination of metrics provides a comprehensive characterization of the controller’s dynamic and steady-state behavior.
% =====================================================================
\subsection{Active power reference step response}
% =====================================================================
The closed-loop transient response is evaluated by applying a step change in the power reference from 300~MW to 600~MW. Two scenarios are considered: nominal conditions and a perturbed case with $L_r = 0.1L_{r0}$ and $R_r = 0.2R_{r0}$.

\begin{figure}[htbp]
    \centering
    \subfloat[\label{fig:StepResponseComp_PMMC1}]{%
        \includegraphics[width=\columnwidth]{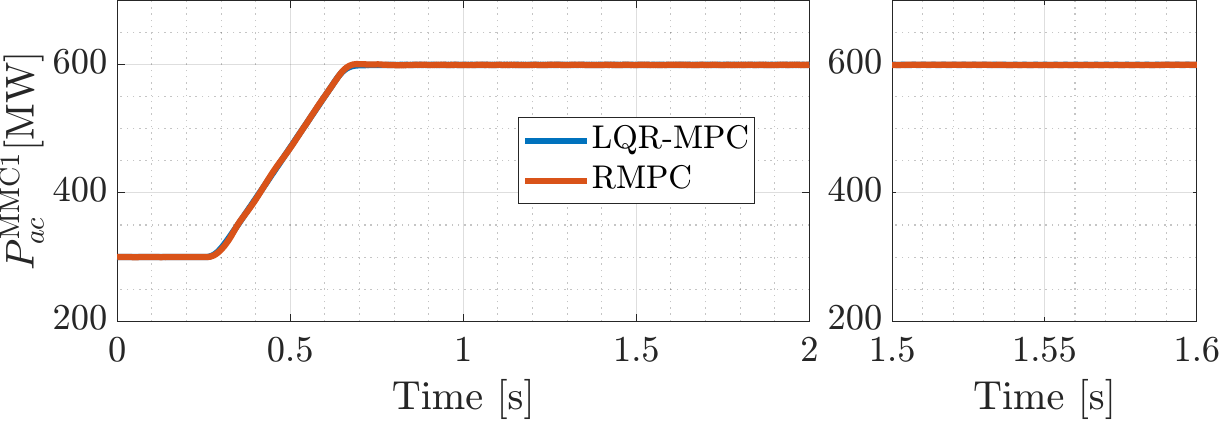}} \\
    \subfloat[\label{fig:StepResponseComp_QMMC1}]{%
        \includegraphics[width=\columnwidth]{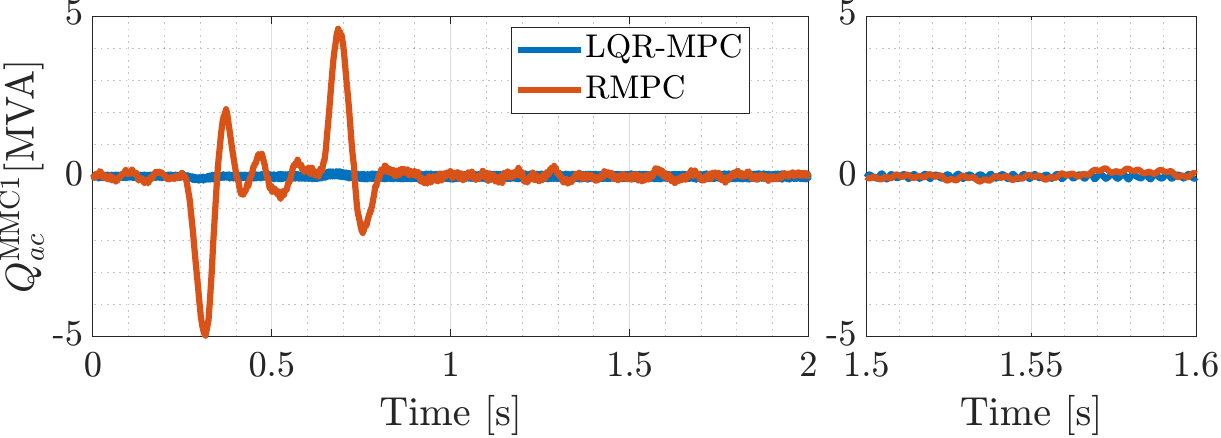}}  
    \caption{(a) Active power $P_{\text{MMC1}}$ and (b) reactive power $Q_{\text{MMC1}}$, trajectories of nominal MMC.1 controlled by LQR-MPC (blue) and RMPC (red).}
    \label{fig:StepResponseComp_All_MMC1}
\end{figure}

\begin{figure}[htbp]
    \centering
    \subfloat[\label{fig:StepResponseComp_IABC1_LQR}]{%
        \includegraphics[width=\columnwidth]{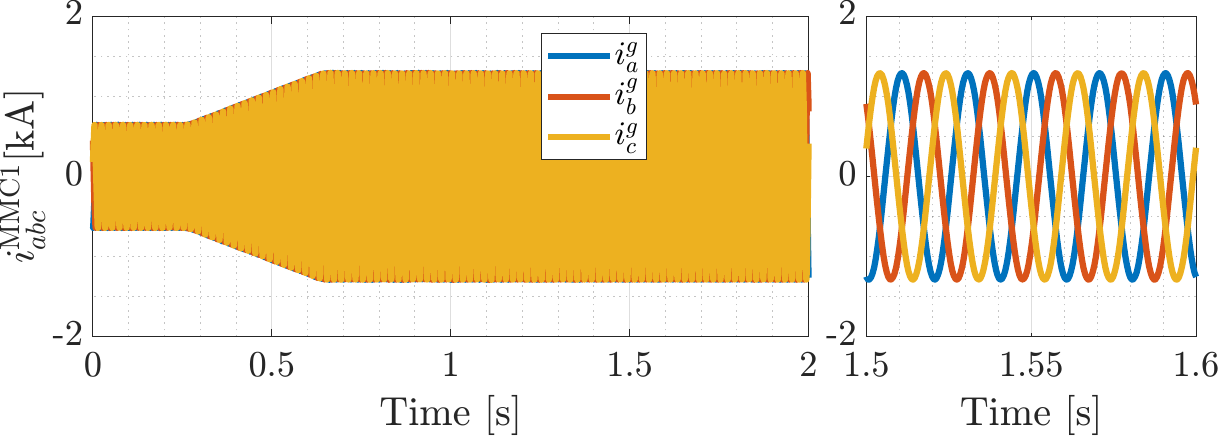}} \\
    \subfloat[\label{fig:StepResponseComp_IABC1_RMPC}]{%
        \includegraphics[width=\columnwidth]{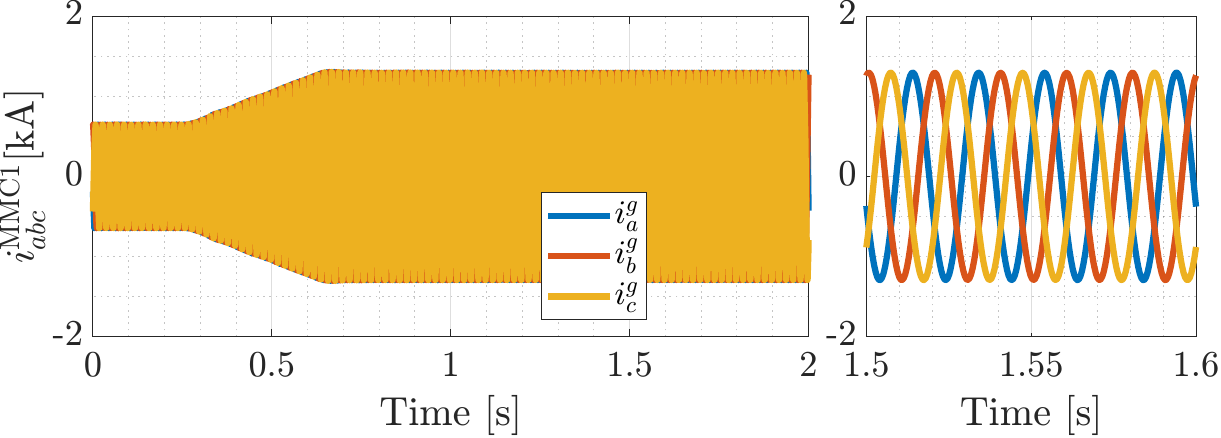}} 
    \caption{AC grid 1 current $i_{g1}$ trajectories of nominal MMC.1 controlled by (a) LQR-MPC and (b) RMPC.}
    \label{fig:StepResponseComp_IABC1_All_RMPC}
\end{figure}

Under nominal conditions, \cref{fig:StepResponseComp_PMMC1} shows the active power trajectories measured at the secondary side of TF.1 for both LQR-MPC and RMPC controllers. 
The responses are nearly identical, exhibiting accurate reference tracking with overlapping trajectories. 
\cref{fig:StepResponseComp_QMMC1} complements this analysis by presenting the reactive power behavior. 
In this case, the RMPC response exhibits slightly higher oscillations and a longer settling time; however, the settling time remains within the 1.5~s range reported in the literature. 
Importantly, as shown in \cref{fig:StepResponseComp_IABC1_All_RMPC}, the reactive power transient does not adversely affect current waveforms or power quality, and both controllers maintain acceptable performance.

\begin{figure}[htbp]
    \centering
    \subfloat[\label{fig:StepResponseComp_Out_PMMC1}]{%
        \includegraphics[width=\columnwidth]{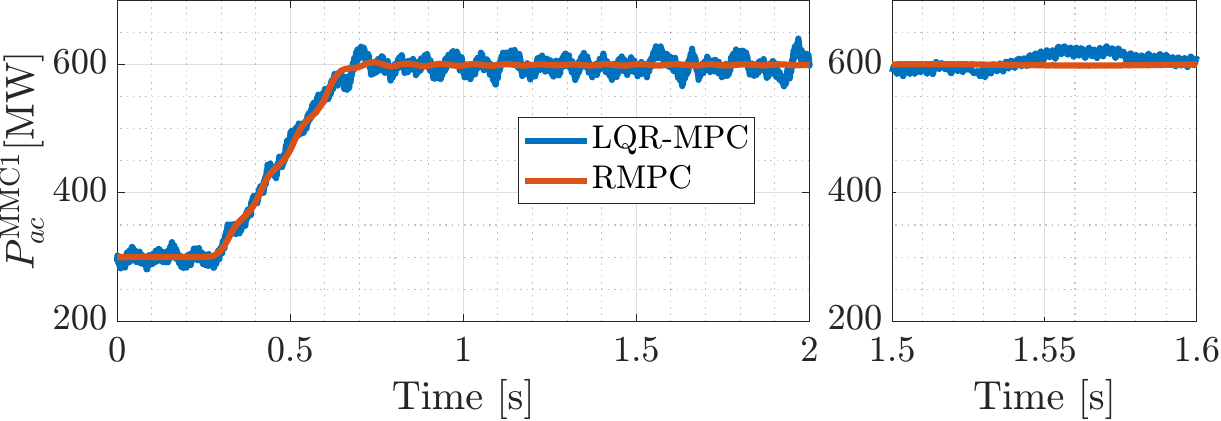}} \\
    \subfloat[\label{fig:StepResponseComp_Out_QMMC1}]{%
        \includegraphics[width=\columnwidth]{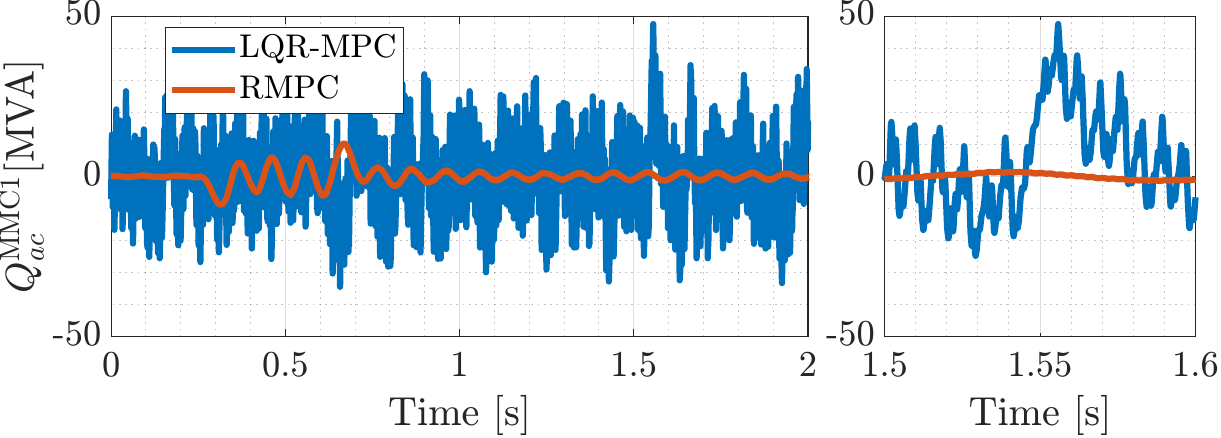}}
    \caption{(a) Active power $P_{\text{MMC1}}$ and (b) reactive power $Q_{\text{MMC1}}$ trajectories of MMC.1 with $L_r = 0.1L_{r0}$ and $R_r = 0.1R_{r0}$ controlled by LQR-MPC (blue) and RMPC (red).}
    \label{fig:StepResponseComp_Out_All_MMC1}
\end{figure}

\begin{figure}[!htbp]
    \centering
    \subfloat[\label{fig:StepResponseComp_Out_IABC1_LQR}]{%
        \includegraphics[width=\columnwidth]{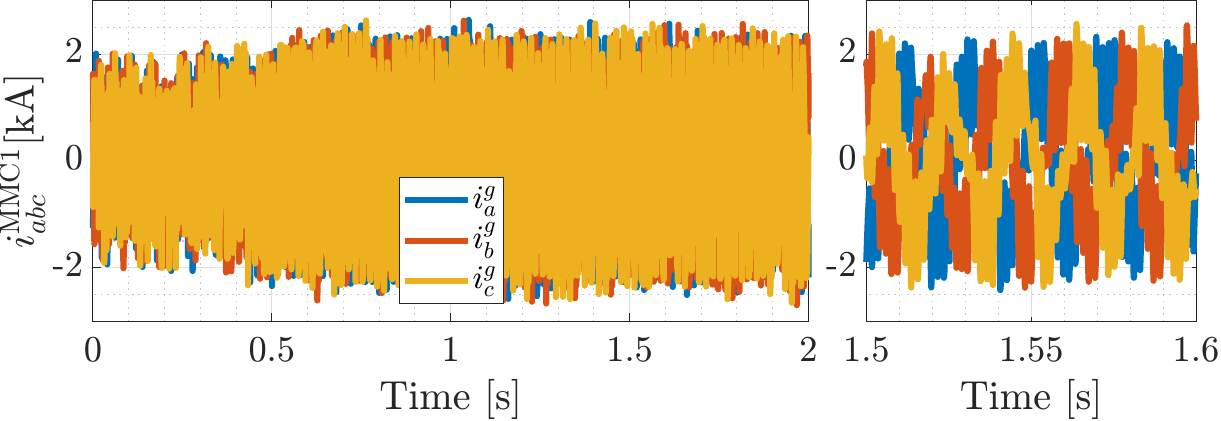}} \\
    \subfloat[\label{fig:StepResponseComp_Out_IABC1_RMPC}]{%
        \includegraphics[width=\columnwidth]{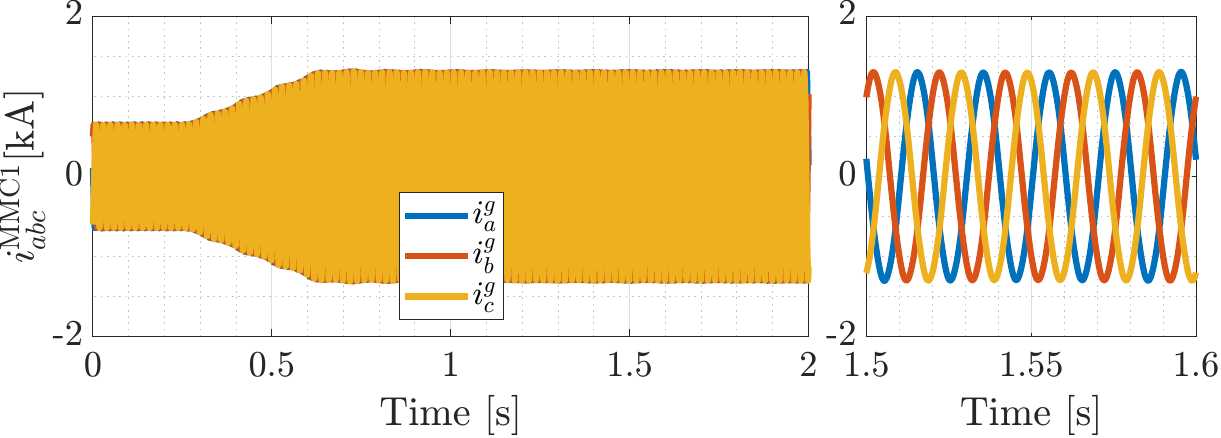}}  
    \caption{AC grid 1 current $i_{g1}$ trajectories of MMC.1 with $L_r = 0.1L_{r0}$ and $R_r = 0.1R_{r0}$ controlled by (a) LQR-MPC and (b) RMPC.}
    \label{fig:StepResponseComp_Out_All_IABC1}
\end{figure}

\begin{figure}[!htbp]
    \centering
    \subfloat[\label{fig:StepResponseComp_Out_VDC2}]{%
        \includegraphics[width=\columnwidth]{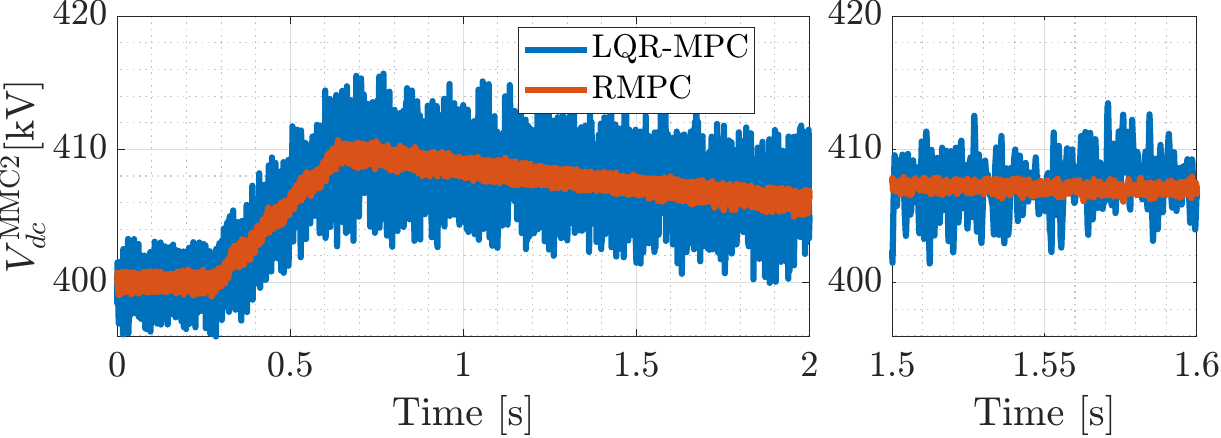}} \\
    \subfloat[\label{fig:StepResponseComp_Out_iDC}]{%
        \includegraphics[width=\columnwidth]{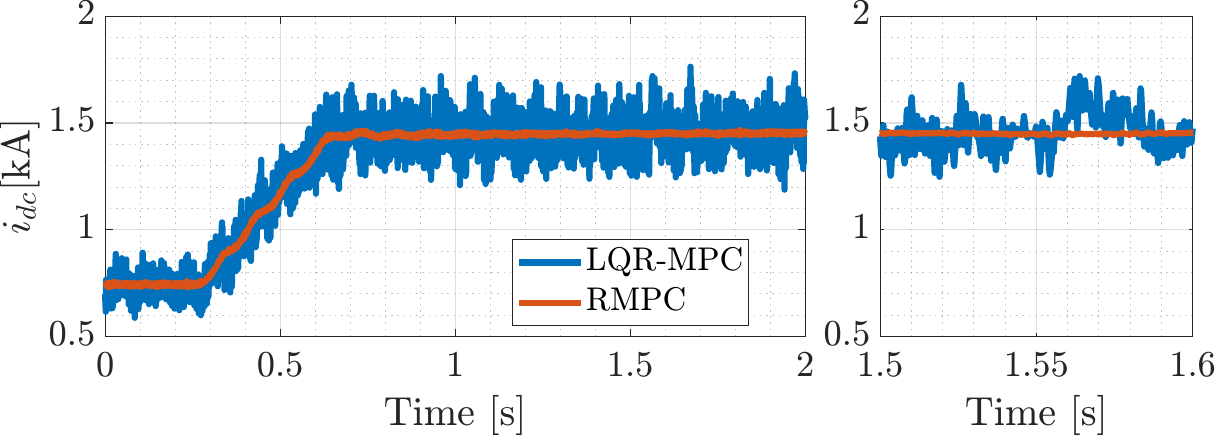}}
    \caption{DC-link (a) current $i_{dc}$ and (b) voltage $V_{dc}$ trajectories of MMC.1 with $L_r = 0.1L_{r0}$ and $R_r = 0.1R_{r0}$ controlled by LQR-MPC (blue) and RMPC (red).}
    \label{fig:StepResponseComp_Out_All_iDC}
\end{figure}

The results for the scenario with modeling mismatches are presented in \cref{fig:StepResponseComp_Out_All_MMC1,fig:StepResponseComp_Out_All_IABC1,fig:StepResponseComp_Out_All_iDC}. 
In this case, the performance of the two controllers differs significantly.
Although eigenvalue analysis in the verification stage could suggests strong robustness of the LQR-MPC tuning, EMT simulations reveal the presence of unmodeled uncertainties that are not captured at that stage. 
While system stability under LQR-MPC is not compromised, transient performance cannot be guaranteed or reliably predicted \emph{a priori} with LQR-based tuning. 
The observed oscillations therefore indicate degraded performance rather than instability.

In contrast, RMPC demonstrates enhanced robustness and consistent stability under parameter variations. This distinction is clearly illustrated in \cref{fig:StepResponseComp_Out_All_IABC1}, where grid current trajectories under LQR-MPC exhibit significant and inadmissible harmonic content, whereas RMPC maintains acceptable waveform quality. 
Finally, the DC-link dynamics (current and voltage), depicted in \cref{fig:StepResponseComp_Out_All_iDC} exhibit comparatively poorer performance with LQR-MPC, showing that uncertainties in the OC dynamics significantly degrade the DC dynamics, despite the decoupling observed in \eqref{eq:discreteTAugementedTimeSSMMC}.

% =====================================================================
\subsection{Frequency-domain RTDS result}
% =====================================================================
% \textcolor{red}{Robust stability margins is analyzed through bode analysis admittance-based transfer function.}

\begin{figure}[htbp]
    \centering
    \includegraphics[width=\linewidth]{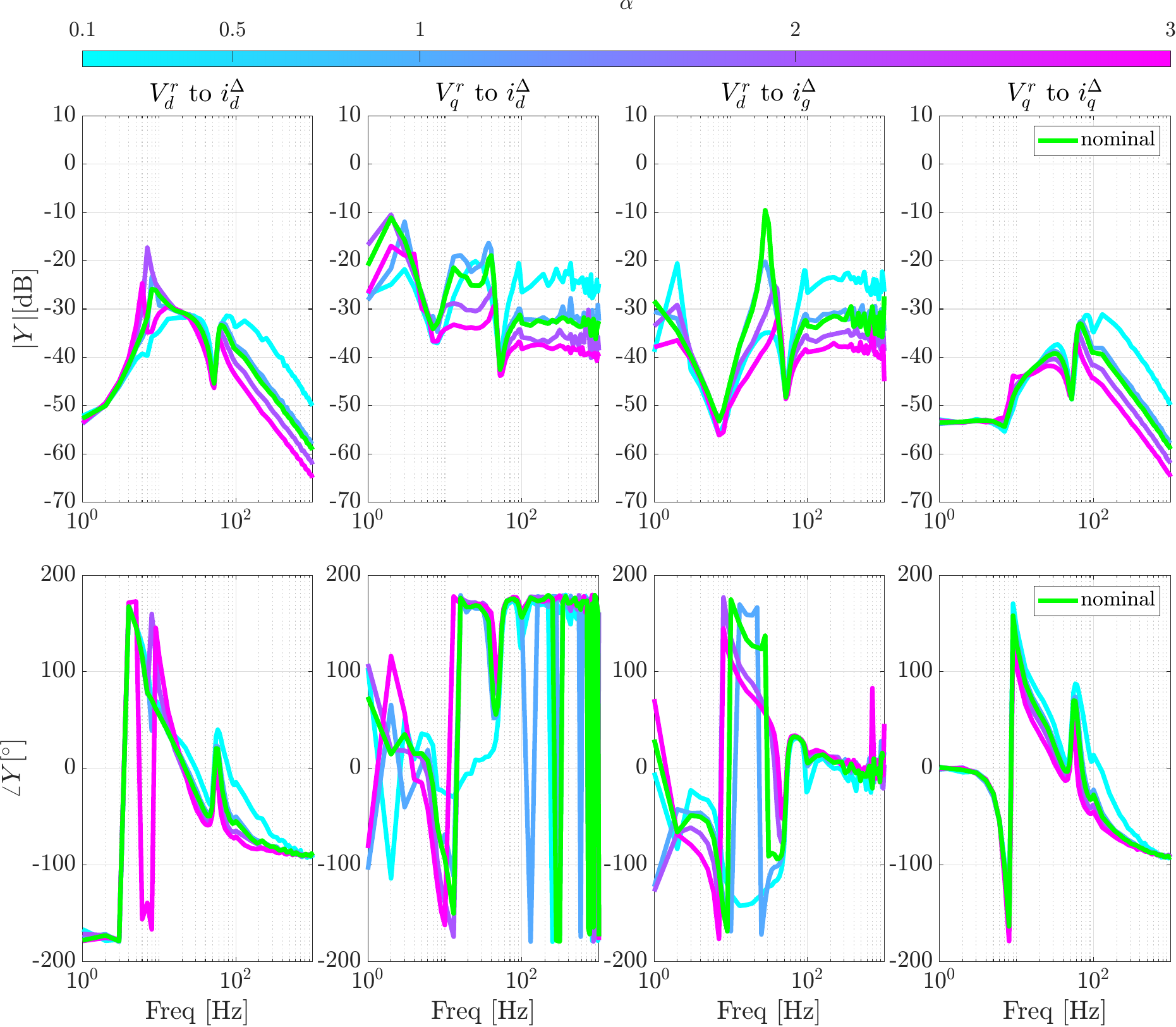}
    \caption{Comparison of AC-side admittance matrix of MMC controlled with RMPC for different \( L_{r} = \alpha L_{r0} \) and \(  R_{r} = \alpha R_{r0} \) and nominal conditions plotted in green.}
    \label{fig:ImpedanceMatrixUncertOC_RMPC}
\end{figure}

\cref{fig:ImpedanceMatrixUncertOC_RMPC} presents the Bode response of the AC-side admittance matrix of MMC.1. The frequency scan is obtained by perturbing the nominal AC grid voltage in the $abc$ frame, with power references set to $P_{\text{MMC1}}=300$~MW and $Q_{\text{MMC1}}=0$~MVA.

The $dd$ and $qq$ responses show attenuation of grid voltage perturbations across the entire frequency range. 
Anti-resonant peaks at 50 Hz confirm the controller's effective regulation of AC components. 
Resonant peaks around 7 Hz are attributed to grid impedance dynamics and DC-link interactions that are not captured in the uncertain model, indicating reduced low-frequency damping. 
This characteristic can be mitigated by relaxing input soft constraints to increase controller bandwidth.
The cross-coupling terms ($dq$ and $qd$) are also attenuated, exhibiting anti-resonant behavior at 7~Hz and 50~Hz. These results are consistent with the verification-stage analysis. 
The phase margin is effectively infinite, while the maximum resonance peak is approximately $-10$~dB, indicating a gain margin exceeding the recommended 6~dB~\footnote{This is an estimation since exact margins are difficult to determine for MIMO systems}.

\begin{figure}[htbp]
    \centering
    \includegraphics[width=\linewidth]{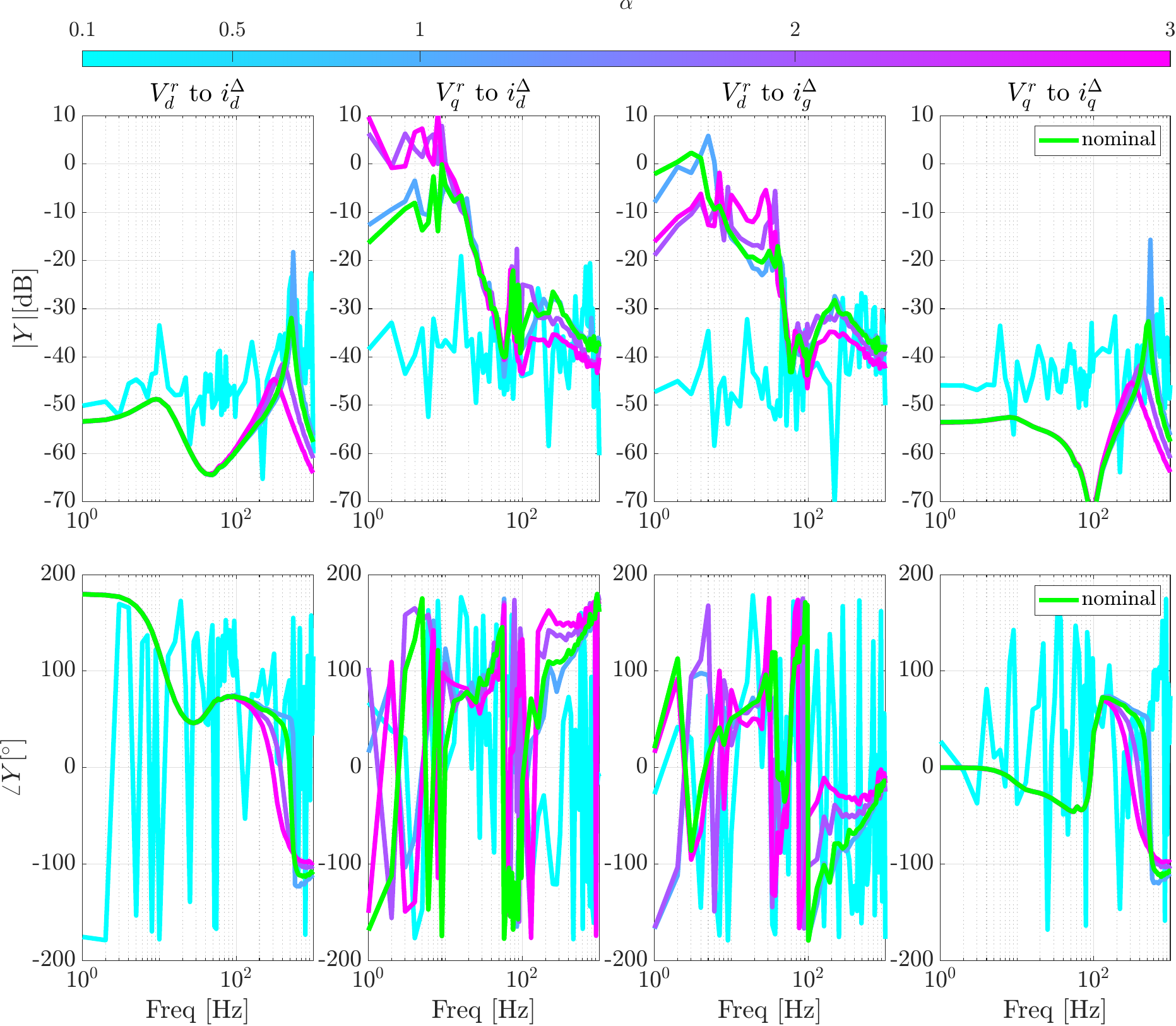}
    \caption{Comparison of AC-side admittance matrix of MMC controlled with LQR-MPC for different \( L_{r} = \alpha L_{r0} \) and \(  R_{r} = \alpha R_{r0} \) and nominal conditions plotted in green.}
    \label{fig:ImpedanceMatrixUncertOC_LQRv3}
\end{figure}

A contrasting behavior is observed for the admittance matrix of the closed-loop system under LQR--MPC control (see \cref{fig:ImpedanceMatrixUncertOC_LQRv3}). 
In all cases, performance degrades as $\alpha$ decreases.
As expected, LQR--MPC exhibits a higher control bandwidth, particularly evident in the nominal $dd$ component, where low-frequency resonant peaks are not observed.
However, this increased bandwidth comes at the expense of higher sensitivity at elevated frequencies, as indicated by pronounced resonant peaks around 520~Hz. 
Moreover, the cross-coupling terms exhibit significant gain at low frequencies, especially under perturbed conditions, where grid voltage disturbances are insufficiently attenuated. 
Although the phase margin remains positive, it is consistently lower than that achieved with RMPC, indicating reduced robustness.

\subsection{Total harmonics distortion assessment}

To assess the power quality achieved by the proposed control tuning, the total harmonic distortion (THD) of the AC grid current is evaluated under steady-state conditions for a power reference of $P^{\text{MMC1}}_{ac}=300$~MW and varying uncertainty realizations defined by the parameter $\alpha$.
At this operating point, the nominal current amplitude at 50~Hz is approximately 0.64~kA.

\begin{figure}[!htbp]
    \centering
    \subfloat[\label{fig:SteadyStateComp_THD_LQR}]{%
        \includegraphics[width=0.9\columnwidth]{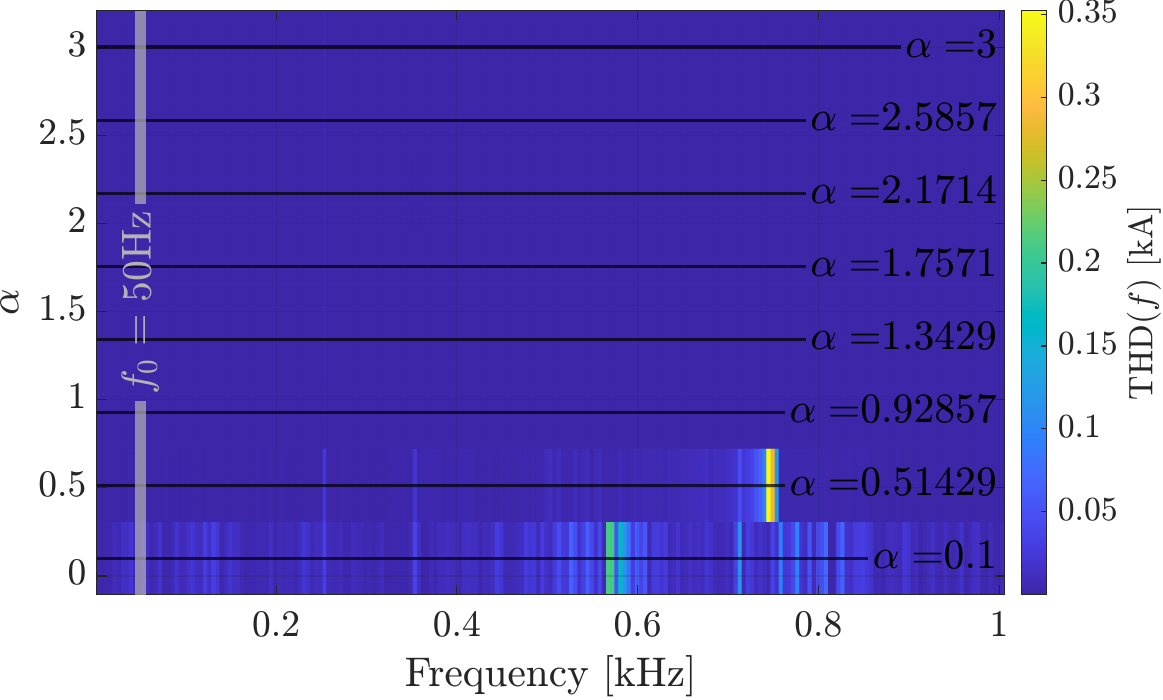}} \\
    \subfloat[\label{fig:SteadyStateComp_THD_RMPC}]{%
        \includegraphics[width=0.9\columnwidth]{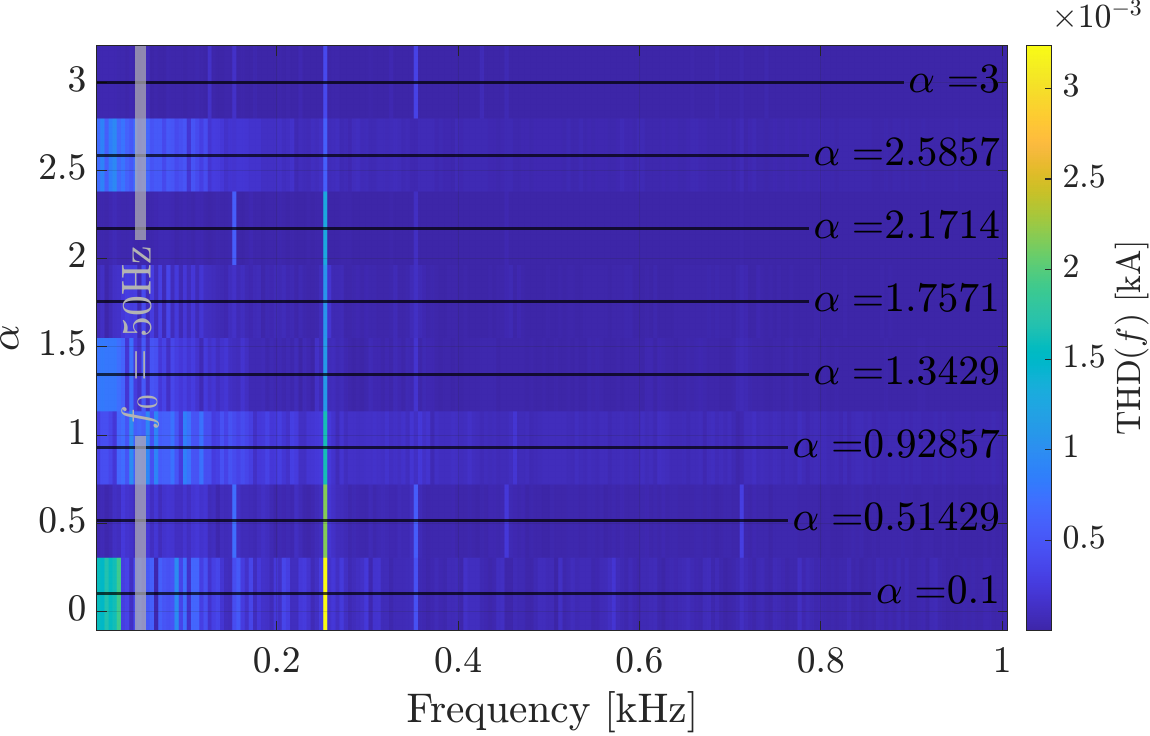}}
    \caption{Heatmap of THD components of AC grid~1 current $i_{g1}$ across frequency and uncertainty realizations $\alpha$, comparing (a) LQR--MPC and (b) RMPC.}
    \label{fig:SteadyStateComp_THD_All_RMPC}
\end{figure}

\cref{fig:SteadyStateComp_THD_LQR} shows the relationship between $\alpha$ and the THD spectrum\footnote{In this case, the component corresponding to the fundamental frequency is excluded from the plot.} obtained via FFT when LQR-MPC is employed. 
For low values of $\alpha$, significant harmonic distortion is observed, with components reaching up to 0.35~kA (approximately 50\% of the nominal current amplitude). For higher $\alpha$, THD levels remain within limits recommended by TSOs.

In contrast, \cref{fig:SteadyStateComp_THD_RMPC} (bottom) presents the THD spectrum under RMPC control across the same uncertainty set. A marked performance improvement is observed, with harmonic components not exceeding 4~A, corresponding to less than 0.3\% of the nominal current amplitude. 
This performance is consistently maintained across all uncertainty realizations.

\section{Conclusion}
This paper presented an automatic tuning method for robust MPC in MMC-based HVDC systems. 
The approach computes MPC weighting matrices via a matching-control framework enhanced with an LMI formulation, reducing conservatism while preserving robustness guarantees. 
The resulting RMPC inherits these properties without increasing online computational complexity, enabling real-time implementation.
The method was validated using simulations on RTDS\textsuperscript{\textregistered}, demonstrating consistent performance under parametric uncertainty and nonlinear conditions. Compared to LQR-based tuning, the proposed approach guarantees closed-loop and impedance-based stability, while directly enforcing phase and gain margin requirements specified by TSOs.
Across all scenarios, the RMPC achieves faster damping, reduced high-frequency oscillations, and lower overcurrent levels. These results confirm a scalable and computationally efficient framework for practical robust MPC implementation in HVDC systems.

\appendices
\section{Controller Gains}
\label{sec_appx:Gains}
% \subsection{RMPC Parameters}
The RMPC parameters are  $K^* =  \operatorname{diag}(K^{\Delta *},K^{\Sigma *})$, $Q^* =  \operatorname{diag}(K^{\Delta *},Q^{\Sigma *})$, $P^* =  \operatorname{diag}(P^{\Delta *},P^{\Sigma *})$ and $R^* = 0.01\textbf{I}_4$ with 
\begin{equation*}
    \begin{aligned}
        K^{\Delta*} &= \begin{bmatrix} -0.1707 & 0.0011 & -0.0009 & -0.0002 \\
        -0.0011 & -0.1707 & 0.0002 & -0.0009\end{bmatrix},\\        
        K^{\Sigma*} &= \begin{bmatrix} 0.1088 & 0.0016 & 0.0008 & -0.0004 \\
        -0.0016 & 0.1088 & 0.0004 & 0.0008\end{bmatrix},      
    \end{aligned}
\end{equation*}
\begin{equation*}
    \begin{aligned}
         Q^{\Delta*} &= \begin{bmatrix} 0.7086 & -0.0007 & -0.0000 & -0.0031 \\
        -0.0007 & 0.7063 & 0.0031 & 0.0000 \\
        -0.0000 & 0.0031 & 0.0000 & 0.0000 \\
        -0.0031 & 0.0000 & 0.0000 & 0.0000\end{bmatrix}1e^{-4}, \\        
        Q^{\Sigma*} &= \begin{bmatrix} 0.8073 & -0.0006 & -0.0000 & 0.0042 \\
        -0.0006 & 0.8057 & -0.0042 & -0.0000 \\
        -0.0000 & -0.0042 & 0.0000 & -0.0000 \\
        0.0042 & -0.0000 & -0.0000 & 0.0000\end{bmatrix}1e^{-5}, 
    \end{aligned}
\end{equation*}
\begin{equation*}
    \begin{aligned}
        % R^* &= ,\\ 
        P^{\Delta*} &= \begin{bmatrix} 0.0282 & -0.0000 & 0.0001 & 0.0001 \\
        -0.0000 & 0.0282 & -0.0001 & 0.0001 \\
        0.0001 & -0.0001 & 0.0000 & 0.0000 \\
        0.0001 & 0.0001 & 0.0000 & 0.0000\end{bmatrix},\\
        P^{\Sigma*} &= \begin{bmatrix} 0.0107 & 0.0000 & 0.0000 & -0.0001 \\
        0.0000 & 0.0107 & 0.0001 & 0.0000 \\
        0.0000 & 0.0001 & 0.0000 & 0.0000 \\
        -0.0001 & 0.0000 & 0.0000 & 0.0000\end{bmatrix}.
    \end{aligned}
\end{equation*}
The LQR-MPC parameters are  $K_{\text{LQR}} =  \operatorname{diag}(K^{\Delta}_{\text{LQR}},K^{\Sigma }_{\text{LQR}})$,  and $P_{\text{LQR}} =  \operatorname{diag}(R^{\Delta}_{\text{LQR}},R^{\Sigma}_{\text{LQR}})$ with 
\begin{equation*}
    \begin{aligned}
        K^{\Delta}_{LQR} &= \begin{bmatrix} -2.1097 & 0.0166 & -0.1466 & -0.0165 \\
        -0.0166 & -2.1097 & 0.0165 & -0.1466\end{bmatrix} \\
        K^{\Sigma}_{LQR} &= \begin{bmatrix}1.5882 & 0.0249 & 0.1423 & -0.0247 \\
        -0.0249 & 1.5882 & 0.0247 & 0.1423\end{bmatrix} \\
    \end{aligned}
\end{equation*}
\begin{equation*}
    \begin{aligned}        
        P^{\Delta}_{LQR} &=  \begin{bmatrix} 686.1588 & -0.0000 & 47.5883 & 5.7483 \\
        -0.0000 & 686.1588 & -5.7483 & 47.5883 \\
        47.5883 & -5.7483 & 7.2180 & -0.0000 \\
        5.7483 & 47.5883 & -0.0000 & 7.2180\end{bmatrix}\\
        P^{\Sigma}_{LQR} &=  \begin{bmatrix}305.0561 & 0.0000 & 27.2046 & -5.1727 \\
        0.0000 & 305.0561 & 5.1727 & 27.2046 \\
        27.2046 & 5.1727 & 5.5814 & 0.0000 \\
        -5.1727 & 27.2046 & 0.0000 & 5.5814\end{bmatrix}
    \end{aligned}
\end{equation*}
The alternative LQR parameters are $K^2_{\text{LQR}} =  \operatorname{diag}(K^{\Delta 2}_{\text{LQR}},K^{\Sigma 2  }_{\text{LQR}})$ with 
\begin{equation*}
    \begin{aligned}
        K^{\Delta 2}_{LQR} &= \begin{bmatrix} -15.0318 & 0.1181 & -8.6551 & -0.0790 \\
        -0.1181 & -15.0318 & 0.0790 & -8.6551\end{bmatrix}\\
        K^{\Sigma 2}_{LQR} &= \begin{bmatrix} 9.2875 & 0.1459 & 5.5785 & -0.0933 \\
        -0.1459 & 9.2875 & 0.0933 & 5.5785\end{bmatrix}
    \end{aligned}
\end{equation*}

\bibliographystyle{IEEEtran}
\bibliography{references}

\end{document}